\documentclass[12pt,a4paper]{article}

\usepackage{epsfig}
\usepackage{amsmath,amsfonts,amssymb}
\usepackage{cite}
\usepackage{colortbl}
\usepackage{pifont}
\usepackage{authblk}
\usepackage{soul}

\def \mn{\mu\nu{\rm SSM}}

\providecommand{\openone}{\leavevmode\hbox{\small1\kern-3.8pt\normalsize1}}

\title{Exotic diboson $Z'$ decays in the U$\mu\nu$SSM}

\author[a]{J.~A.~Aguilar-Saavedra\thanks{jaas@ugr.es}}
\author[b]{I.~Lara\thanks{inaki.lara@fuw.edu.pl}}
\author[c,d]{D.~E.~L\'opez-Fogliani\thanks{daniel.lopez@df.uba.ar}}
\author[e,f]{C.~Mu\~noz\thanks{c.munoz@uam.es}} 

\affil[a]{Departamento de F\'{\i}sica Te\'orica y del Cosmos,  Universidad de Granada,  E-18071 Granada, Spain}
 \affil[b] {Faculty of Physics, University of Warsaw, Pasteura 5, 02-093 Warsaw, Poland}
\affil[c]{Instituto de F\'isica de Buenos Aires UBA \& CONICET, Departamento de F\'isica, Facultad de Ciencia Exactas y Naturales, Universidad de Buenos Aires, 1428 Buenos Aires, Argentina}
\affil[d]{Pontificia Universidad Católica Argentina, 
Av. Alicia Moreau de Justo 1500, 
1107~Buenos~Aires, Argentina}
\affil[e]{Departamento de F\'{\i}sica Te\'{o}rica, Universidad Aut\'{o}noma de Madrid (UAM), Campus de Cantoblanco, 28049 Madrid, Spain}
\affil[f]{Instituto de F\'{\i}sica Te\'{o}rica (IFT) UAM-CSIC,  Campus de Cantoblanco, 28049 Madrid, Spain}

\begin{document}

\maketitle

\begin{abstract}
Searches for new leptophobic resonances at high energy colliders usually target their decay modes into pairs of light quarks, top quarks, or standard model bosons. Additional decay modes may also be present, producing signatures to which current searches are not sensitive. We investigate the performance of generic searches that look for resonances decaying into two large-radius jets. As benchmark for our analysis we use a supersymmetric $\text{U}(1)'$ extension of the Standard Model, the so-called U$\mu\nu$SSM, where all the SM decay modes of the $Z'$ boson take place, plus additional (cascade) decays into new scalars. The generic searches use a generic multi-pronged jet tagger and take advantage of the presence of $b$ quarks in the large-radius jets, and are sensitive to all these $Z'$ decay modes (except into light quarks) at once. For couplings that are well below current experimental constraints, these generic searches are sensitive at the $3\sigma-4\sigma$ level with Run 2 LHC data.
\end{abstract}

\section{Introduction}

Realistic extensions of the Standard Model (SM) involve several new particles in addition to the already discovered ones, which likely appear at different mass scales. Their collider signals, arguably, are often not captured by the simplified scenario framework in which most searches at the Large Hadron Collider (LHC) are performed. The variety of possible signals makes dedicated searches unmanageable, therefore, a generic approach is compulsory.

The simplest scenario where generic searches become very useful is the decay of a very heavy resonance $R$ (at the multi-TeV scale) into two particles at the electroweak scale, which can in turn decay into quarks, or into lighter new particles which ultimately decay into quarks. The signature of such cascade decay is a pair of massive jets, each of them corresponding to one of the decay products of $R$. Among the possible signatures with this topology, one can have
\begin{enumerate}
\item Dibosons: $R \to VV$, $V=W,Z$, or $Vh$, with $h$ the Higgs boson. Those signatures are profusely investigated at the LHC~\cite{Sirunyan:2019jbg,Aaboud:2018eoy,Aad:2019fbh,Sirunyan:2021bzu}.
\item Exotic dibosons: $R \to B_1 B_2$, with $B_i$ new scalars decaying into quarks (for masses below the $t \bar t$ threshold the dominant decay channel is expected to be $B_i \to b \bar b$).
\item Exotic multibosons: $R \to S_1 S_2$, with $S_i$ additional scalars decaying $S_i \to B_1 B_2$. Such signals have been dubbed as `stealth'~\cite{Aguilar-Saavedra:2017zuc} because the four-pronged jets from the boosted $S$ decay are difficult to pinpoint using two-pronged taggers.
\end{enumerate}
Searches for dibosons are not sensitive to exotic dibosons if the masses of the latter $m_{B_i}$ are far from the $W,Z,h$ masses, so that the jet mass windows used in the event selection (e.g. typically $60-100$ GeV for $W$ and $Z$ bosons) do not contain a sizeable fraction of the signal. The same applies to stealth bosons $S_i$, with the extra penalty that the taggers for two-pronged jets may further suppress the signal. It is not difficult to write down models where diboson signals are absent, for example with a new $Z'$ resonance that does not decay into $WW$ because the $Z-Z'$ mixing vanishes, and instead can decay into new scalars. A minimal implementation has been presented in Ref.~\cite{Aguilar-Saavedra:2019adu}. 

In order to search for all these signatures at once, one needs a strategy that generalises the usual diboson resonance searches. Such strategy relies on (i) a generic tagger, not only for two-pronged jets, but rather for any type of multi-pronged jets; (ii) $b$ tagging of subjets inside the large-radius jets, to take advantage of the expected presence of $b$ quarks in the cascade decays.
With this strategy, searches are sensitive to resonances $R$ of any spin, and decay products that can be either scalars or vector bosons, because the generic tagger is so designed.

As framework to study these signals we consider the recently proposed
U$\mu\nu$SSM~\cite{Aguilar-Saavedra:2021qbv},\footnote{See also Refs.~\cite{Fidalgo:2011tm,Lozano:2018esg} for similar scenarios.} which is a $U(1)'$ extension of the `$\mu$ from $\nu$' supersymmetric standard 
model ($\mu\nu$SSM)~\cite{LopezFogliani:2005yw}
(for a recent review, see Ref.~\cite{Lopez-Fogliani:2020gzo}). In such a framework the three decay modes (diboson, and exotic di/multiboson) are present. This approach is the opposite to the one adopted in Ref.~\cite{Aguilar-Saavedra:2019adu}. There, a minimal model was written down in which the usual diboson decays $Z' \to WW$ are absent, so that generic searches sensitive to exotic dibosons and multibosons are compulsory. Here, we focus on a model in which all the three types of bosonic decays are present, to assess the sensitivity of a generic search in that situation and see, with a concrete example, how the different $Z'$ bosonic decay modes contribute to the combined sensitivity. Other models with a $Z'$ boson may have one or more of the decays described above.

In U$\mu\nu$SSM scenarios, the presence of $R$-parity violating couplings involving right-handed (RH) neutrino
superfields, $\hat \nu^c$, solves simultaneously the 
$\mu$ problem~\cite{Kim:1983dt} (for a recent review, see Ref.~\cite{Bae:2019dgg})
of the minimal supersymmetric standard model (MSSM)
(for reviews, see e.g. Refs.~\cite{Nilles:1983ge,Haber:1984rc,Martin:1997ns}) and the
$\nu$ problem being able to reproduce the neutrino data~\cite{LopezFogliani:2005yw,Escudero:2008jg,Ghosh:2008yh,Bartl:2009an,Fidalgo:2009dm,Ghosh:2010zi,Fidalgo:2011tm,Aguilar-Saavedra:2021qbv}.
In the superpotential of these scenarios, in addition to Yukawa couplings for neutrinos 
$Y^{\nu} \, \hat H_u\hat L\hat \nu^c$,  the couplings $\lambda\, \hat \nu^c\hat H_d \hat H_u$ are allowed generating an effective $\mu$-term when  the RH sneutrinos 
develop electroweak-scale vacuum expectation values (VEVs),
$ \langle {\widetilde \nu}_{R}\rangle = v_{R} / \sqrt{2} \sim 1$ TeV.

The U$\mn$ models built in Ref.~\cite{Aguilar-Saavedra:2021qbv} are constrained by anomaly cancellation conditions and have the attractive properties of forbidding baryon-number-violating operators as well as explicit Majorana masses and $\mu$ terms, and avoiding potential domain wall problems. 
Besides, the extra $U(1)'$ gauge symmetry provides the RH neutrinos with a non-vanishing charge, avoiding the uneasy situation from the theoretical viewpoint of being the only fields with no quantum numbers under the gauge group.
The presence of several right sneutrinos as well as of the two Higgs doublets $H_d$ and $H_u$, provides a large spectrum of scalar particles which could give rise to the aforementioned multiboson signals. 

In particular, assuming three families of RH neutrino superfields $\hat\nu^c_i$,
$i=1,2,3$, there are eight (seven) neutral scalar (pseudoscalar) states from the mixing between Higgses and sneutrinos. However, the three left sneutrinos are almost decoupled, and upon diagonalisation we are left in the scalar sector 
with the SM-like Higgs, the heavy doublet-like neutral Higgs and three CP-even singlet-like states, and in the pseudoscalar sector with the heavy doublet-like pseudoscalar and three 
CP-odd singlet-like states.
Therefore, decay channels into dijets from a pair production at the LHC of 
CP-even and CP-odd right sneutrinos via the $Z'$, are viable in the U$\mn$. 
In this work, we will carry out a detailed analysis of this signal.

The paper is organised as follows. 
Sec.~\ref{sec:2} will be devoted to the discussion of the U$\mu\nu$SSM benchmarks we use to illustrate the production of exotic multiboson signals.
In Sec.~\ref{sec:3} we examine the sensitivity to such signals with generic searches. Finally, our conclusions and a discussion of possible generalisations are left for Sec.~\ref{sec:4}.

\section{U$\mu \nu$SSM benchmarks}
\label{sec:2}

In this work we consider scenario 1 of Ref.~\cite{Aguilar-Saavedra:2021qbv} in which the $Z'$ boson is leptophobic. This is the case of interest for resonance searches in hadronic final states: otherwise, the $Z'$ decay into lepton pairs gives much cleaner and distinctive signals. There are other possibilities for leptophobic $Z'$ bosons in U$\mu\nu$SSM models, and the results are quantitatively quite similar. The $\text{U}(1)'$ charges of the relevant fields are collected in Table~\ref{tab:hyp}.

The superpotential of the U$\mu \nu$SSM and interactions of the $Z'$ boson are summarised in Ref.~\cite{Aguilar-Saavedra:2021qbv}, and we refer the reader to that work for details. We restrict ourselves here to the aspects of the model most directly related to our present analysis.

\begin{table}[t!]
\begin{center}
\begin{tabular}{|c|c|c|c|}
\hline
$z(L) = 0$ & $z(e^c) = 0 $ & $z(\nu^c) = \frac{1}{4}$ & $z(H_d) = 0$ \\
\hline
$z(Q) = \frac{1}{36}$ & $ z(u^c) = \frac{2}{9} $ & $z(d^c) = - \frac{1}{36}$ & $z(H_u)=- \frac{1}{4}$ \\
\hline
\end{tabular}
\end{center}
\caption{$\text{U}(1)'$ charges of the SM matter 
within the U$\mu \nu$SSM scenario considered.}
\label{tab:hyp}
\end{table}

\begin{figure}[t]
\begin{center}
\includegraphics[height=5.5cm,clip=]{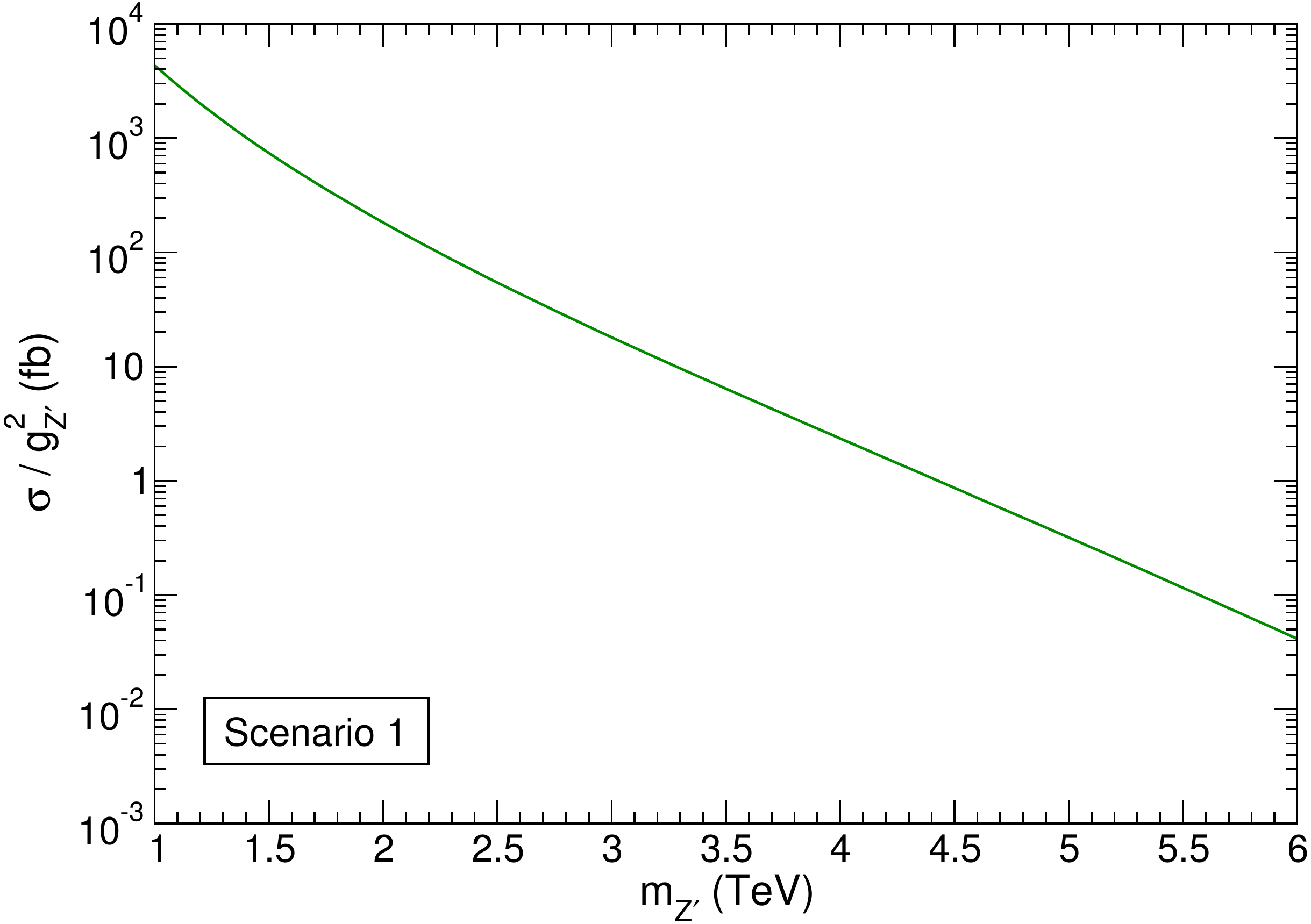} 
\caption{$Z'$ production cross section in scenario 1 of Ref.~\cite{Aguilar-Saavedra:2021qbv}.}
\label{fig:xsec}
\end{center}
\end{figure}

The total $Z'$ production cross section for the scenario considered is presented in Fig.~\ref{fig:xsec} as a function of its mass. The $Z'$ partial width into fermion pairs is
\begin{align}
\Gamma(Z' \to f \bar f) & = \frac{N_c g_{Z'}^2}{24\pi} m_{Z'} \left( 1- 4 \frac{m_{f}^2}{m_{Z'}^2} \right)^{\frac{1}{2}}
\left\{ [z(f)^2 + z(f^c)^2] \left[ 1- \frac{m_{f}^2}{m_{Z'}^2} \right] \right. \notag \\
& \left. - 6 z(f) z(f^c) \frac{m_{f}^2}{m_{Z'}^2} \right\} \,,
\end{align}
with $g_{Z'}$ the $U(1)'$ gauge coupling, $N_c$ the number of colours and $m_f$ the fermion mass. Notice that the $Z'$ boson predominantly decays into up-type quarks, given the $U(1)'$ charges in Table~\ref{tab:hyp},
where $z(f)\ (-z(f^c))$ are the charges for the left (right) chiral fermions. 
The $Z'$ partial width into SM boson pairs is
\begin{align}
\Gamma(Z' \to W^+ W^-) & = \frac{g_{Z'}^2 }{48 \pi} 
 [z(H_u) \sin^2 \beta - z(H_d) \cos^2 \beta ]^2 m_{Z'}
\left( 1- 4 \frac{m_{W}^2}{m_{Z'}^2} \right)^{3/2} \notag \\
& \times \left( 1 + 20  \frac{m_{W}^2}{m_{Z'}^2} + 12 \frac{m_{W}^4}{m_{Z'}^4} \right) 
\notag \\
\Gamma(Z' \to Zh) & = \frac{g_{Z'}^2}{48 \pi} 
[z(H_u) \sin^2 \beta - z(H_d) \cos^2 \beta ]^2\
\frac{\lambda^{1/2}(m_{Z'}^2,m_h^2,m_Z^2)}{m_{Z'}} \notag \\
& \times \left( 1+10 \frac{m_{Z}^2}{m_{Z'}^2} - 2 \frac{m_{h}^2}{m_{Z'}^2} + \frac{m_{Z}^4}{m_{Z'}^4}
 + \frac{m_{h}^4}{m_{Z'}^4} - 2 \frac{m_Z^2 m_h^2}{m_{Z'}^4} \right) \,,
\end{align}
where  $\tan\beta\equiv v_u/v_d$ is the usual ratio of neutral scalar VEVs, e.g. in the MSSM, and we have assumed the alignment limit in the latter channel. We have defined the usual kinematical function
\begin{equation}
\lambda(x,y,z) = x^2 + y^2 + z^2 - 2xy -2xz -2yz \,.
\end{equation}
Notice that for the leptophobic scenario considered $z(H_d) = 0$. These two partial widths are nearly equal for $m_{Z'} \gg m_{W,Z,h}$.

In addition, we have decays into scalar pairs. The interaction of the new gauge boson with neutral scalar fields, generically denoted as $\phi$, is
\begin{equation}
\mathcal{L} = - i g_{Z'} z(\phi) \,   \phi^* \overleftrightarrow{\partial^\mu}  \phi \,  Z_\mu^\prime \,,
\label{ec:lphi}
\end{equation}
with $z(\phi)$ the corresponding $U(1)'$ charge. If one writes $\phi = (\phi_R + i \phi_I)/\sqrt 2$, the interaction reads
\begin{equation}
\mathcal{L} = g_{Z'} z(\phi) \,   \phi_R \overleftrightarrow{\partial^\mu}  \phi_I \,  Z_\mu^\prime \,,
\label{ec:lphi2}
\end{equation}
connecting a CP-even weak eigenstate $\phi_R$ and a CP-odd one $\phi_I$. 
Assuming CP conservation in the scalar mixing, the mass eigenstates are either CP-even (denoted generically by $H$) or CP-odd (denoted by $A$) and the $Z'$ decays into pairs $H_i A_j$. 
{Considering that additional singlet scalars under the SM gauge group can be present
in U$\mn$ models~\cite{Aguilar-Saavedra:2021qbv}, 
we will take the simplifying assumption that they are heavier than the $Z'$ and 
do not have 
a significant mixing with $H, A$. Therefore, the partial width into pairs of scalar eigenstates $s_1$, $s_2$ with the same $\text{U}(1)'$ charges is}
\begin{eqnarray}
\Gamma(Z' \to s_1 s_2) = \frac{g_{Z'}^2 z(s)^2}{48\pi }  \frac{\lambda^{3/2}(m_{Z'}^2,m_{s_1}^2,m_{s_2}^2)}{m_{Z'}^5} \,.
\label{ec:Gss}
\end{eqnarray}

\begin{figure}[t]
\begin{center}
\begin{tabular}{cc}
\includegraphics[height=5.5cm,clip=]{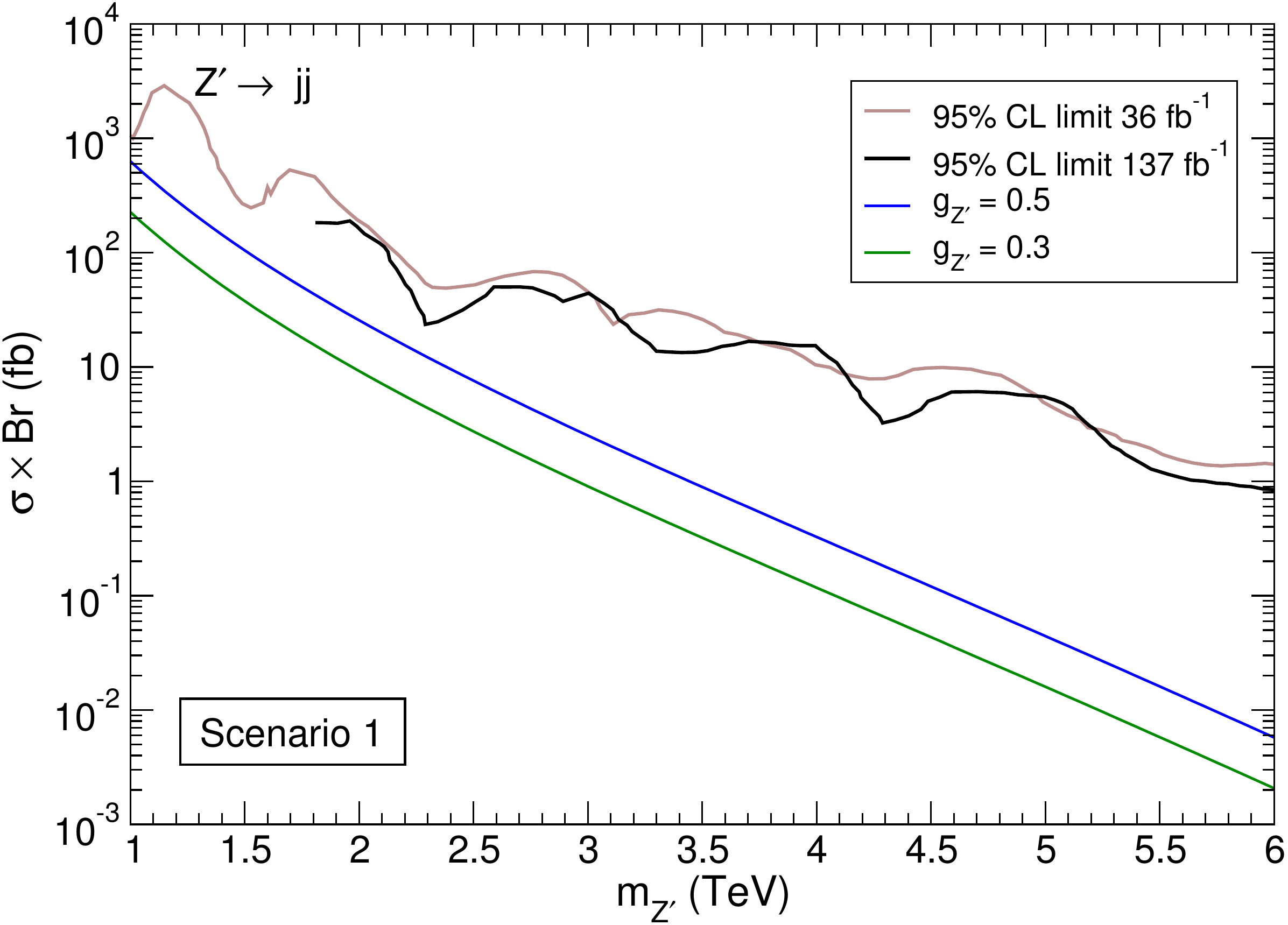} &
\includegraphics[height=5.5cm,clip=]{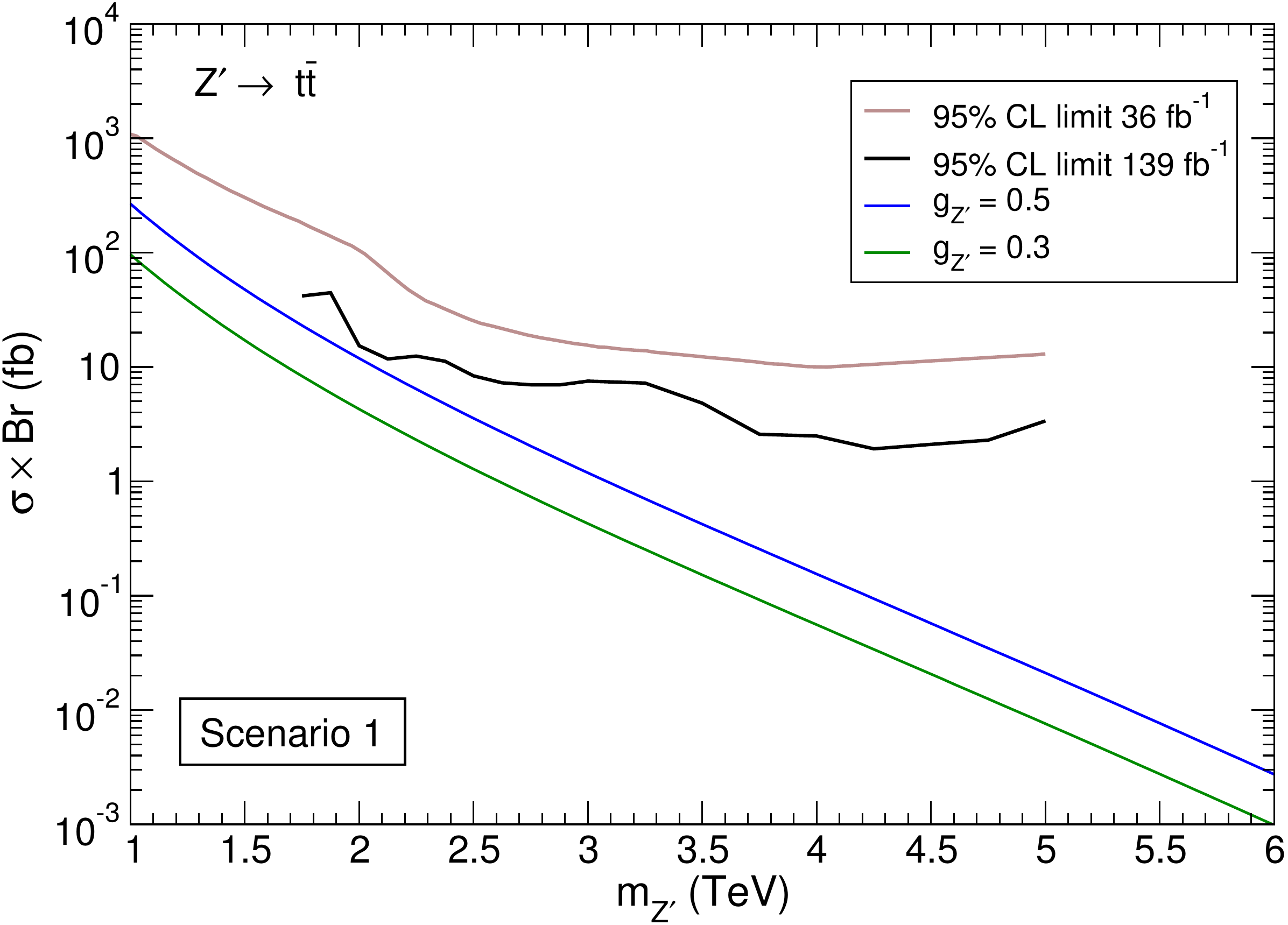}  \\
\includegraphics[height=5.5cm,clip=]{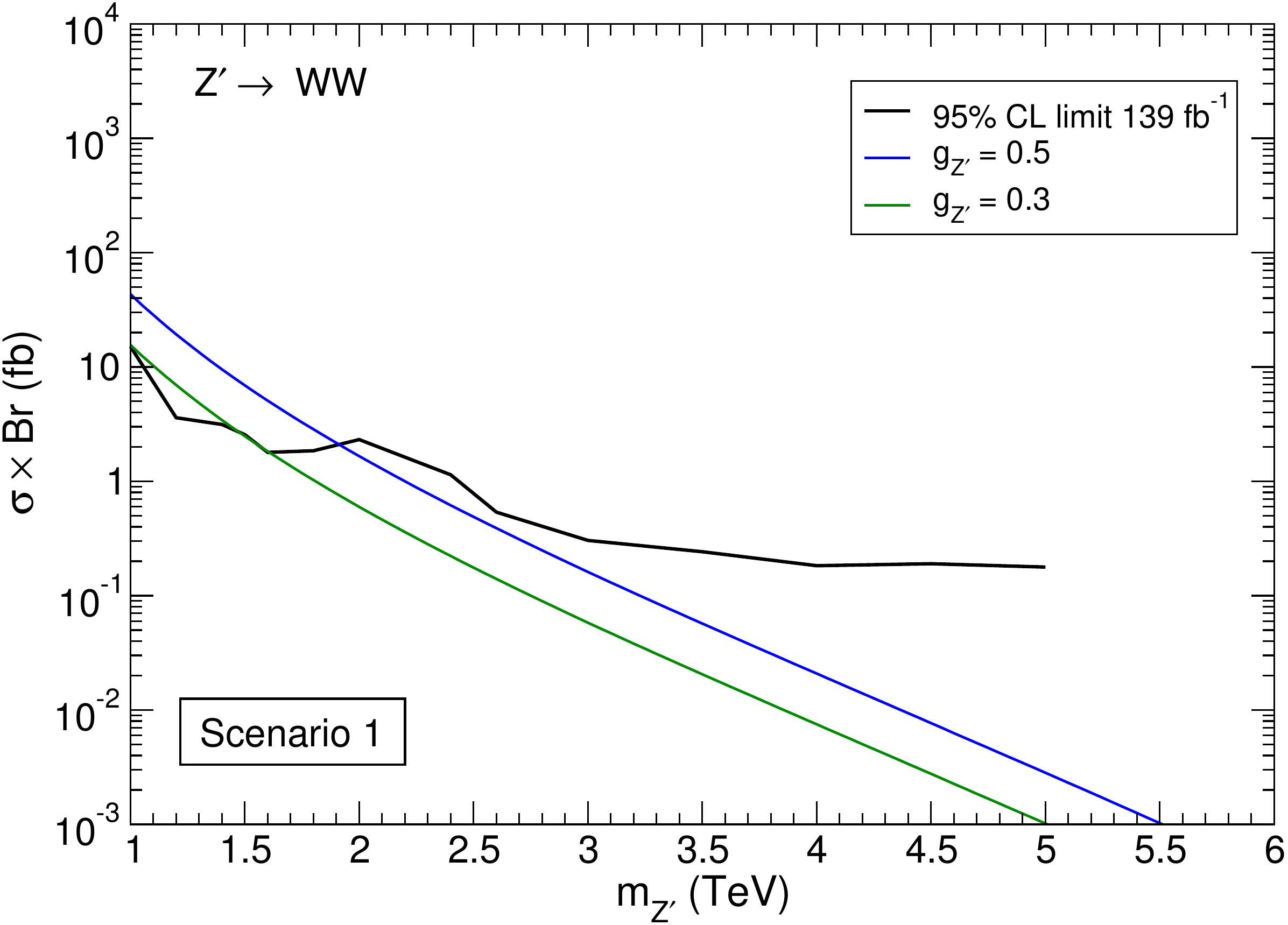}  &
\includegraphics[height=5.5cm,clip=]{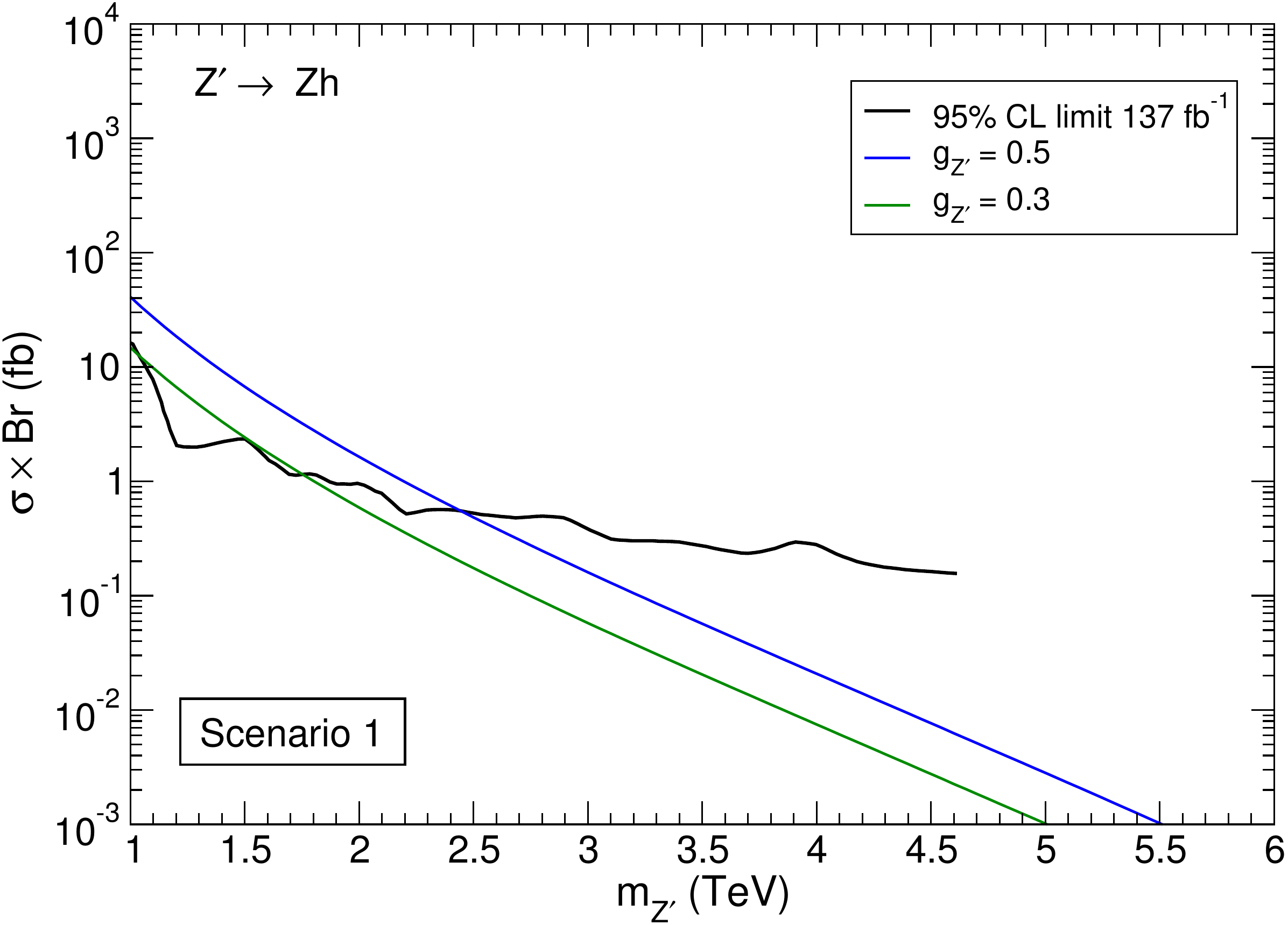} 
\end{tabular}
\caption{Limits on the cross section times branching ratio for the production of $Z'$ bosons arising from searches in several final states: dijets~\cite{Sirunyan:2018xlo,Sirunyan:2019vgj} (top left), $t \bar t$~\cite{Aaboud:2018mjh,Aad:2020kop} (top right), $WW$~\cite{Aad:2019fbh} (bottom left) and $Zh$~\cite{Sirunyan:2021bzu} (bottom right). Together, we show two cross section predictions for $g_{Z'} = 0.3, 0.5$.}
\label{fig:lim}
\end{center}
\end{figure}

The presence of the new decay modes in (\ref{ec:Gss}) slightly decreases the branching ratio into SM final states, and relaxes the current limits with respect to those in Ref.~\cite{Aguilar-Saavedra:2021qbv}. In our numerical benchmarks we will assume these two additional channels: $Z' \to H_2 A_2$, and $Z' \to H_1 A_1$, with $m_{H_i,A_i} \ll m_{Z'}$.
In the U$\mn$ these dibosons are two CP-even and two CP-odd right sneutrinos, {${\widetilde \nu}_{iR}=(H_i+v_R+iA_i) / \sqrt{2}$}.
The current limits from the $Z'$ boson mass and couplings from dijet~\cite{Sirunyan:2018xlo,Sirunyan:2019vgj}, $t \bar t$~\cite{Aaboud:2018mjh,Aad:2020kop}, diboson~\cite{Aad:2019fbh} and $Zh$~\cite{Sirunyan:2021bzu} resonance searches are presented in Fig.~\ref{fig:lim}. We also show the cross section predictions for two $g_{Z'}$ values. As in Ref.~\cite{Aguilar-Saavedra:2021qbv}, we have assumed $\tan \beta = 2$. In the following, we will use two benchmarks,
\begin{itemize}
\item $m_{Z'} = 2$ TeV and $g_{Z'} = 0.3$, with $m_{H_1,A_1} = 80$ GeV, $m_{H_2,A_2} = 200$ GeV.
\item $m_{Z'} = 3$ TeV and $g_{Z'} = 0.5$, with $m_{H_1,A_1} = 96$ GeV, $m_{H_2,A_2} = 240$ GeV.
\end{itemize}
The $Z'$ mass and coupling in both cases are well below the current 95\% confidence level (CL) upper limits. As mentioned, we have assumed CP conservation, so that the neutral scalar mass eigenstates have a definite parity. For simplicity, we also assume that the $(H_i,A_i)$ pairs, $i=1,2$, have the same mass, and that the rest of scalars present in the model are heavier. The $Z'$ is narrow in both scenarios, with $\Gamma_{Z'} = 1.35$ GeV and $\Gamma_{Z'} = 5.6$ GeV, respectively, and the intrinsic $Z'$ width is not relevant for the calculations.

\begin{figure}[t]
\begin{center}
\begin{tabular}{ccc}
\includegraphics[height=3.5cm,clip=]{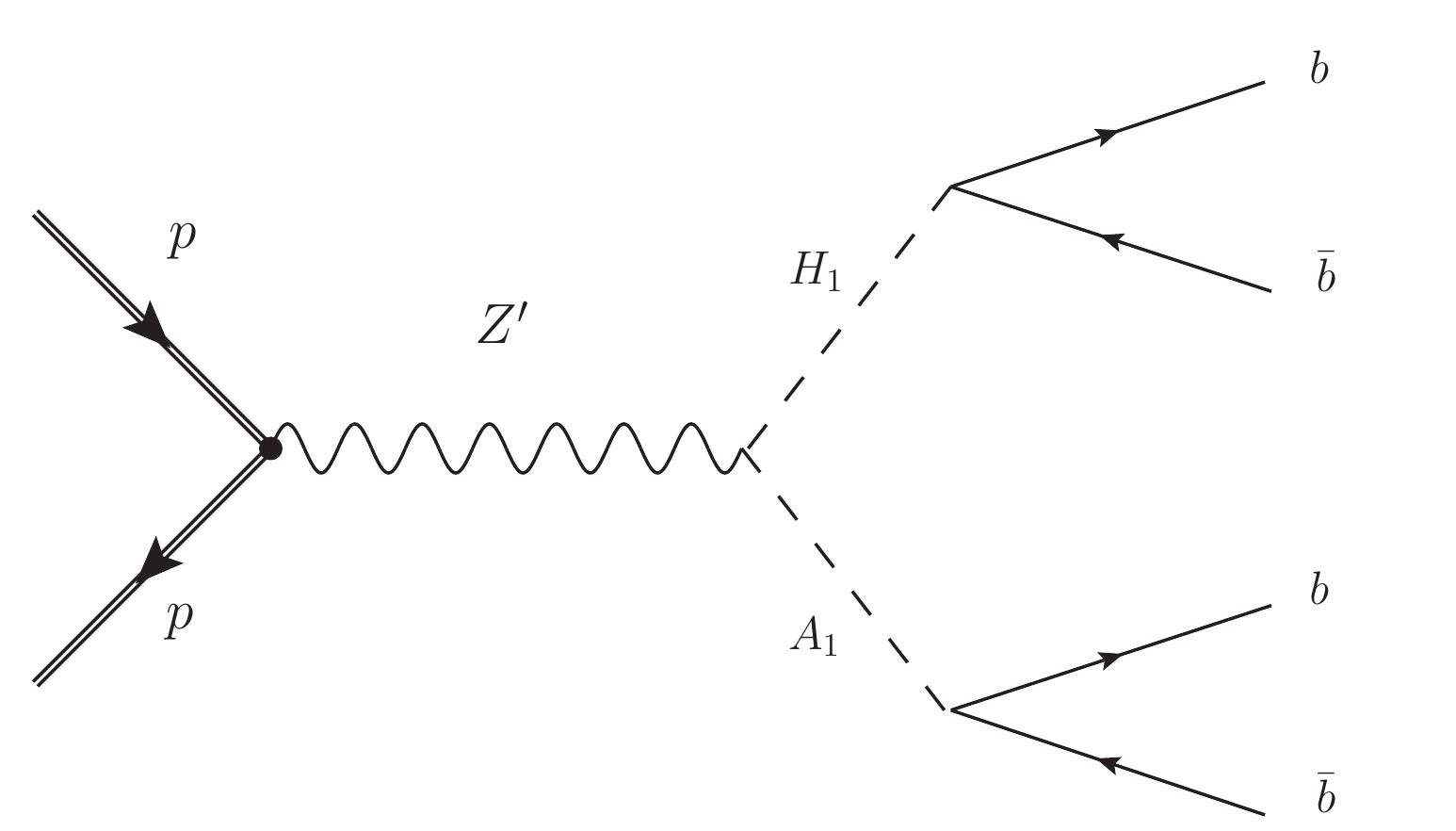} & \quad &
\includegraphics[height=3.5cm,clip=]{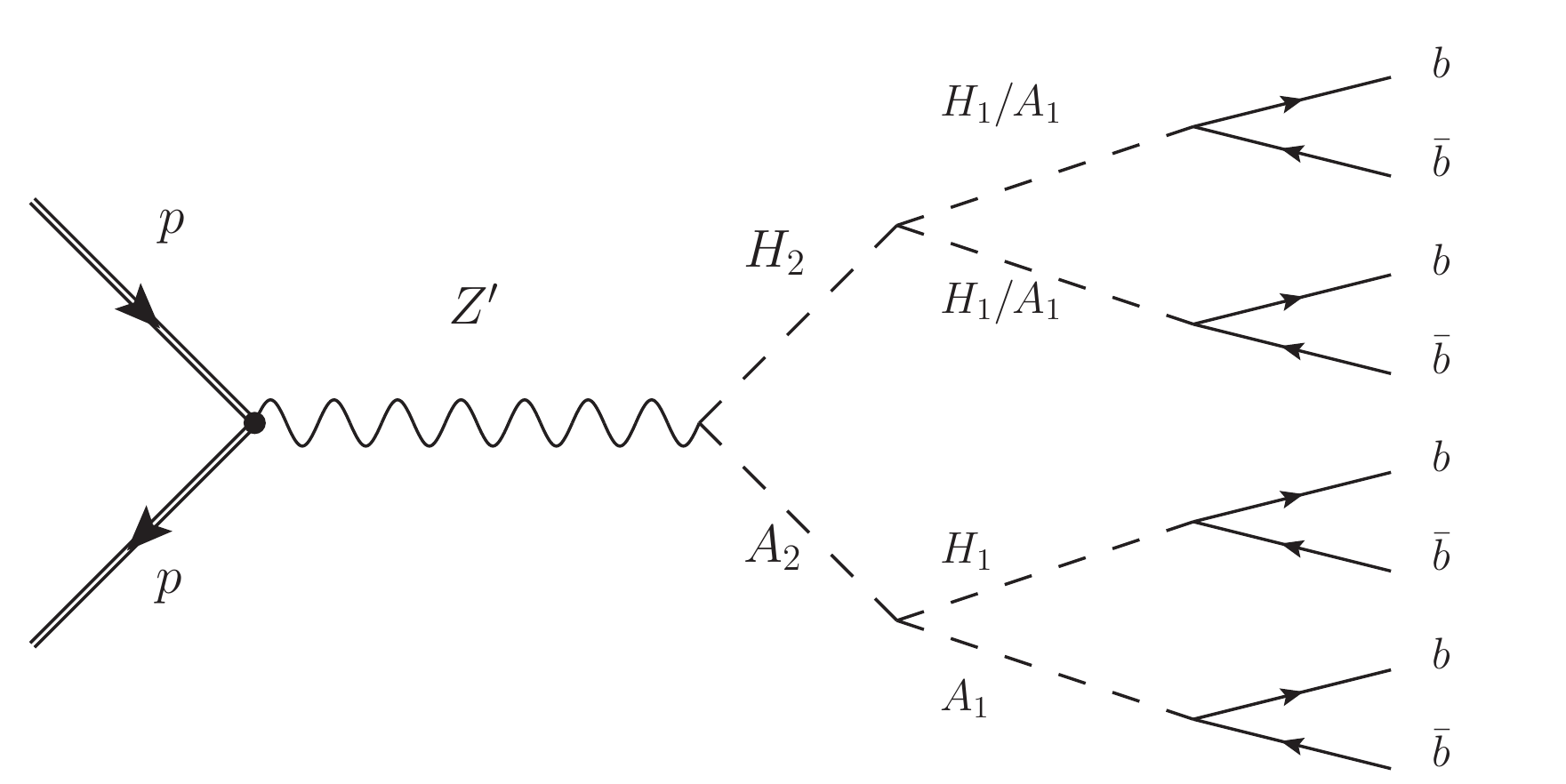}
\end{tabular}
\caption{Cascade decay of the $Z'$ boson into scalars producing exotic diboson and multiboson final states.}
\label{fig:feynman}
\end{center}
\end{figure}

\begin{figure}[t!]
\begin{center}
\begin{tabular}{cc}
\includegraphics[height=4.9cm,clip=]{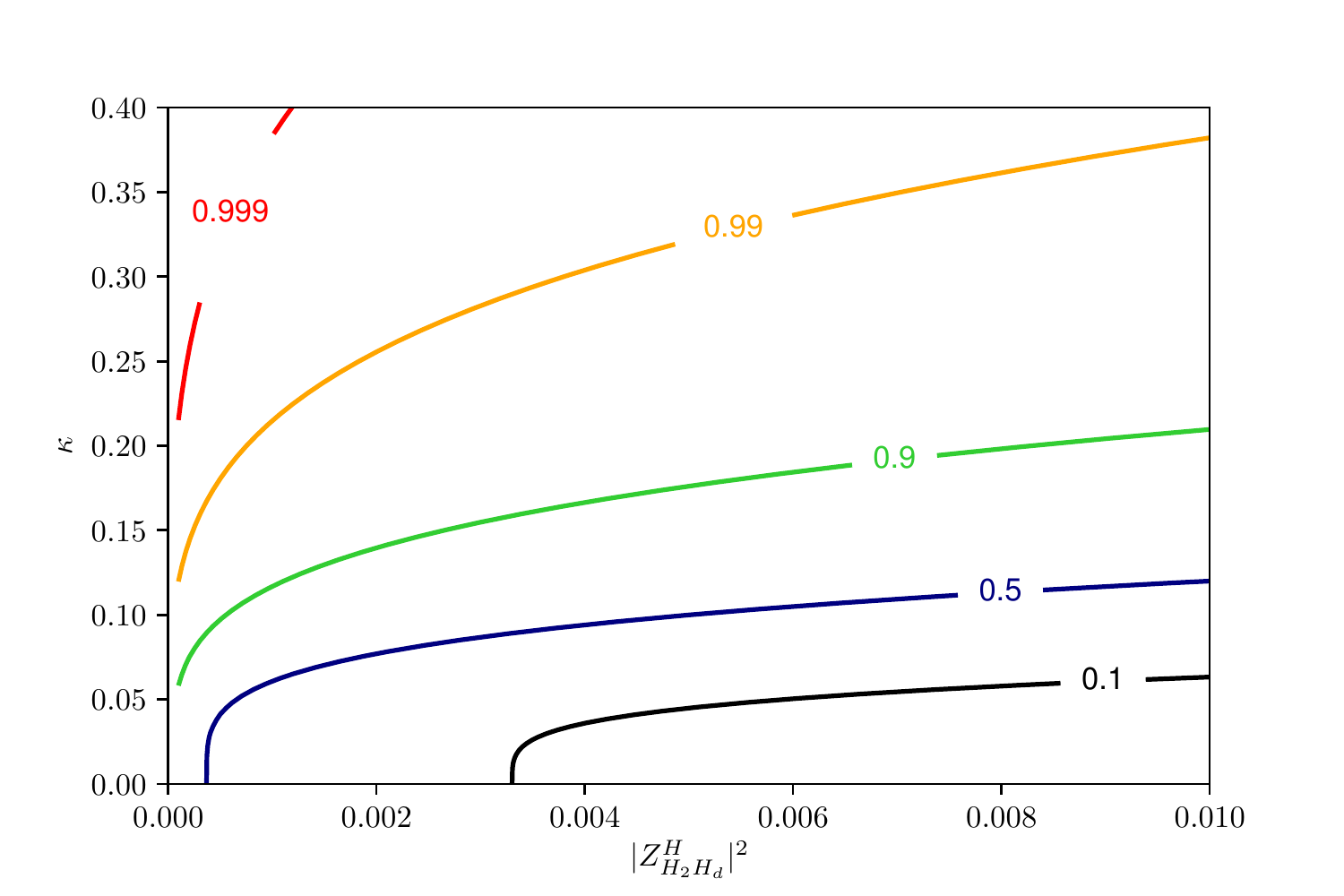} &
\includegraphics[height=4.9cm,clip=]{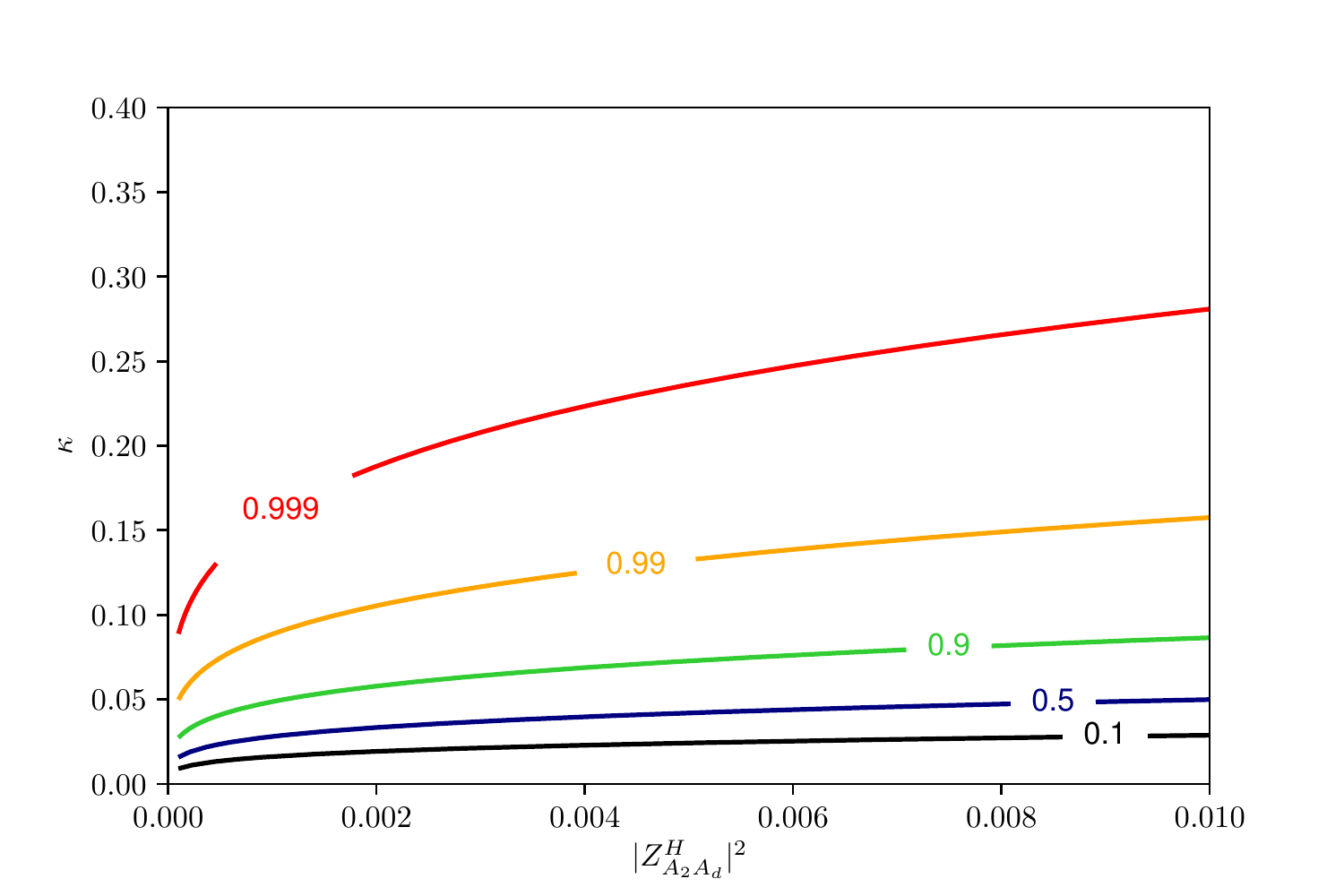}
\end{tabular}
\caption{Left: Lines showing different values of the branching ratio of $H_2$ to $H_1 H_1 + A_1 A_1$, for different values of the coupling $\kappa$ and of
the doublet scalar component of the SM singlet-like scalar $H_2$, $|Z^H_{H_2H_d}|^2$.
The results 
correspond to the benchmark point in the text with $m_{H_2,A_2} = 200$ GeV, $m_{H_1,A_1} = 80$ GeV and $g_{Z'}=0.3$, using
$\tan\beta=2$ and $v_R=1000$ GeV. Right: {The same, but for the decay $A_2 \to A_1 H_1$ and for different values of the doublet pseudoscalar component of the SM singlet-like pseudoscalar $A_2$, $|Z^H_{A_2A_d}|^2$}.}
\label{fig:H2dec}
\end{center}
\end{figure}

The new scalars $H_1, A_1$ are expected to decay predominantly into $b \bar b$ via a small mixing with the {neutral scalar and pseudoscalar Higgses (denoted by $H_d$ and $A_d$ respectively)}, as depicted in Fig.~\ref{fig:feynman} (left). The heavier CP-even scalar $H_2$ can decay into $WW$, $ZZ$ and $b \bar b$ via small mixings with the neutral {scalar} Higgses $H_u$ and $H_d$, but in relevant regions of the parameter space is expected to predominantly decay into $H_1 H_1$ and $A_1 A_1$, as depicted on the right panel. The partial widths for the different modes are
\begin{align}
\Gamma(H_2\to WW) & = \frac{g^2}{64\pi}m_{H_2}\left(\frac{m_{H_2}}{m_W}\right)^2 
\left(
Z^H_{H_2H_d}\cos \beta +
Z^H_{H_2H_u}
\sin\beta \right)^2
\left( 1-4 \frac{m_W^2}{m_{H_2}^2} \right)^{1/2}
\notag \\
& \times
\left(1-4 \frac{m_W^2}{m_{H_2}^2}+12 \frac{m_W^4}{m_{H_2}^4}\right) \,,
\displaybreak
\notag \\[1mm]
\Gamma(H_2\to ZZ) & = \frac{g^2}{128\pi}m_{H_2}\left(\frac{m_{H_2}}{m_W}\right)^2
\left(
Z^H_{H_2H_d}\cos \beta +
Z^H_{H_2H_u}
\sin\beta \right)^2
\left( 1-4 \frac{m_Z^2}{m_{H_2}^2} \right)^{1/2}
\notag \\
& \times
\left(1-4 \frac{m_Z^2}{m_{H_2}^2}+ 12 \frac{m_Z^4}{m_{H_2}^4}\right) \,,
\notag \\[1mm]
\Gamma(H_2\to b \bar b) & = \frac{N_c g^2}{32 \pi} m_{H_2}    \left(\frac{m_b}{m_W}\right)^2 \left(
\frac{Z^H_{H_2H_d}}{\cos \beta} \right)^2 
\left(1-4 \frac{ m_b^2}{m_{H_2}^2}\right)^{3/2} \,,
\notag \\[1mm]
\Gamma(H_2 \to H_1 H_1) & = \frac{1}{64 \pi} m_{H_2} \left(\frac{v_R}{m_{H_2}}\right)^2
\left[{g^2_{Z'}} {z}(\nu^c)^2+2 \kappa^2\right]^2 \left( 1-4\frac{ {m^2_{H_1}}}{{m^2_{H_2}}} \right)^{1/2}\,, 
\notag \\[1mm]
\Gamma(H_2 \to A_1 A_1) & =  \frac{1}{64 \pi}m_{H_2} \left(\frac{v_R}{m_{H_2}}\right)^2 
\left[{g^2_{Z'}} {z}(\nu^c)^2 {-}  2 \kappa^2\right]^2
 \left( 1-4\frac{{m}^2_{A_1}}{{m^2_{H_2}}} \right)^{1/2}\,.
\end{align}
Here $g$ is the $SU(2)$ gauge coupling, $|Z^H_{H_2H_d}|^2$ and $|Z^H_{H_2H_u}|^2$ are the 
{$H_d$ and $H_u$}
components of the SM singlet-like scalar $H_2$, and $\kappa$ is a trilinear coupling {between RH neutrino superfields and an additional superfield $S$, singlet under the SM gauge group,} {$\kappa \hat S\hat \nu^c_i\hat \nu^c_i$, that is allowed to be present in the superpotential dynamically generating Majorana masses for 
RH neutrinos~\cite{Aguilar-Saavedra:2021qbv}}. 
The values of
$|Z^H_{H_2H_d}|^2$ and $|Z^H_{H_2H_u}|^2$ are expected to be small, in the range $10^{-2}-10^{-4}$ (see e.g. Ref.~\cite{Kpatcha:2019qsz}) and we can take them equal for simplicity in our discussion. As we can see in Fig.~\ref{fig:H2dec} {(left)}, for $|Z^H_{H_2H_d}|^2$ in the upper end of the expected range, $H_2$ predominantly decays into two scalars for $\kappa \geq 0.1$, and has branching ratio to $H_1 H_1 + A_1 A_1$ close to unity for $\kappa \geq 0.25$. In the computation of the next section we will take this branching ratio equal to unity. Note that the decays $H_2 \to H_1 h$ are suppressed by the square of the small mixing factors $Z^H$. On the other hand, $H_2 \to hh$ is not suppressed, and is possible if kinematically open. When $h$ decays into $b \bar b$ (which is the dominant decay mode of the 125 GeV Higgs boson) the decays $H_2 \to hh$ produce similar signals to $H_2 \to H_1 H_1 + A_1 A_1$, and for simplicity we consider benchmarks where $H_2 \to hh$ is kinematically forbidden. Higgs decays to final states other than $b \bar b$ also produce massive jets but a dedicated study is beyond the scope of this work.

The heavier CP-odd scalar $A_2$ is {expected to decay into $H_1 A_1$ for analogous reasons}. The partial widths for the possible modes are
\begin{align}
\Gamma({A_2\to b \bar b}) & = 
\frac{N_c g^2}{32 \pi}  
\frac{m_b^2}{m_W^2}m_{A_2}\left(Z^A_{A_2A_d} \tan\beta\right)^2 
\left(1-4\frac{ m_b^2}{m^2_{A_2}}\right)^{\frac{1}{2}} \,,
\notag \\[1mm]
\Gamma({A_2\to H_1A_1}) & =\frac{1}{32\pi}\frac{v_R^2}
{{m^3_{A_2}}} 
\left[
2 \kappa^2\right]^2
\lambda^{1/2}(m_{A_2}^2,m_{A_1}^2,m_{H_1}^2)
 \,,
\end{align}
where $|Z^A_{A_2A_d}|^2$ is the 
small {$A_d$}
component of the SM singlet-like pseudoscalar $A_2$.
{As it can be seen from Fig.~\ref{fig:H2dec} (right), 
the decay $A_2 \to H_1 A_1$ is dominant for $\kappa \geq 0.05$, and has branching ratio near unity for $\kappa \geq 0.1$.}
Overall, in the benchmark points we have considered the decay of the $Z'$ boson into the scalars $H_2$, $A_2$ produces the cascade decays depicted in Fig.~\ref{fig:feynman} (right). Note, however, that longer cascades would in principle be possible. for example if we had considered a large mass differences between $H_i$ and $A_i$ (we have taken $m_{H_i} = m_{A_i}$). In addition,
cascade decays may involve the third pair $H_3$, $A_3$ that we have assumed much heavier. In any case, these two simplified benchmarks depicted in Fig.~\ref{fig:feynman} illustrate the two mechanisms of production of two multi-pronged large radius jets from the $Z'$ decay into new scalars.

\section{Search strategies}
\label{sec:3}

The strategy to search for the dijet signals from multiple cascade decays of the $Z'$ boson into scalars takes advantage of the common features expected:
\begin{itemize}
\item[(1)] sizeable jet masses;
\item[(2)] multi-pronged jet structure;
\item[(3)] multiple $b$ quarks that can be tagged.
\end{itemize}
Because we are not looking for a specific signal, a narrow jet mass window cannot be a priori imposed. Neither a dedicated tagger can be used; rather, a generic tagger for multi-pronged jets~\cite{Aguilar-Saavedra:2017rzt,Aguilar-Saavedra:2020sxp,Aguilar-Saavedra:2020uhm},  or generic anomaly detection methods~\cite{Collins:2018epr,Collins:2019jip,Dillon:2019cqt,Dillon:2020quc,Nachman:2020lpy,Andreassen:2020nkr,Khosa:2020qrz,Heimel:2018mkt,Farina:2018fyg,Hajer:2018kqm,Blance:2019ibf,Amram:2020ykb,Cheng:2020dal} must be used. In order to take advantage of the presence of $b$ quarks, the sample is divided into different categories corresponding to the number of $b$ tags, and the results are subsequently combined.

We perform a simulation including the backgrounds from dijet production ($jj$), $b \bar b$, and $t \bar t$. Event samples are generated with {\scshape MadGraph}~\cite{Alwall:2014hca} in 100 GeV bins starting at $[300,400]$ GeV up to $[2.1,2.2]$ TeV, plus a bin with $p_T \geq 2.2$ TeV. Each dijet sample contains $6 \times 10^5$ events, and each $b \bar b$ and $t \bar t$ sample $10^5$ events. For each background process ($jj$, $b \bar b$, $t \bar t$) there are in total 20 samples in different $p_T$ bins which are combined with weight proportional to the cross section, so as to have good statistics across a wide range of transverse momentum. 
For the signal processes the relevant Lagrangian is implemented in {\scshape Feynrules}~\cite{Alloul:2013bka} and interfaced to {\scshape MadGraph5} using the universal Feynrules output~\cite{Degrande:2011ua}.
Hadronisation and parton showering performed with {\scshape Pythia~8}~\cite{Sjostrand:2007gs} and detector simulation with {\scshape Delphes 3.4}~\cite{deFavereau:2013fsa} using the configuration for the CMS detector. The reconstruction of jets and the analysis of their substructure is done using {\scshape FastJet}~\cite{Cacciari:2011ma}. 

In the analysis we use a collection of `fat' jets with radius $R=0.8$ reconstructed with the anti-$k_T$ algorithm~\cite{Cacciari:2008gp} and groomed with Recursive Soft Drop~\cite{Dreyer:2018tjj}. 
In addition, for $b$ tagging we use a collection of `track jets' of radius $R = 0.2$, reconstructed using only tracks.
This procedure is similar to the one applied for $b$-tagging of boosted $H \to b \bar b$ by the ATLAS Collaboration~\cite{Aad:2020tps}. A fat jet is considered to have a single $b$ tag if there is one (and only one) $b$-tagged track jet within a distance $\Delta R = 0.8$. A double $b$ tag on the fat jet is considered if there are two or more $b$-tagged track jets within a distance $\Delta R = 0.8$. The fat jet is untagged if there are no $b$-tagged tagged track jets within $\Delta R = 0.8$. For the tagging of track jets we use an updated Run 2 parametric efficiency formula~\cite{Sirunyan:2017ezt} in the 50\% efficiency working point. Despite the fact that this parameterisation was not explicitly derived for $R=0.8$ jets, we believe it still a sufficiently good approximation for our purpose, with misidentification probabilities around $0.14$ for charm jets and $0.05$ for light jets at the $p_T$ range of interest.

\begin{figure}[t]
\begin{center}
\includegraphics[height=5.2cm,clip=]{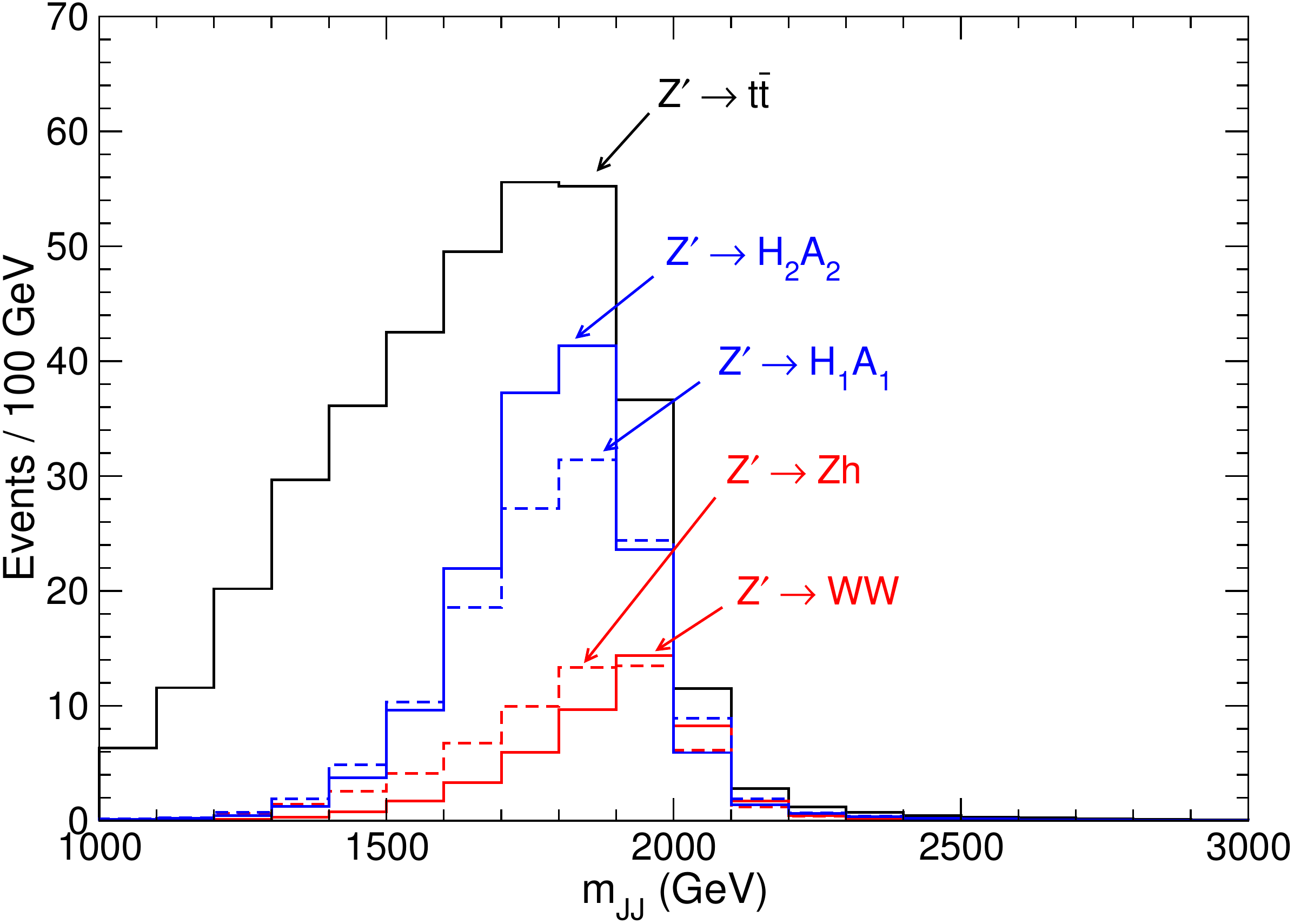} 
\caption{Dijet invariant mass for several $Z'$ decays in the benchmark with $m_{Z'} = 2$ TeV.}
\label{fig:mJJ-Zp}
\end{center}
\end{figure}

As event pre-selection we require the presence of two jets with mass $m_J \geq 50$ GeV, transverse momentum $p_T \geq 400$ GeV, rapidity difference $|\Delta \eta| \leq 1.3$, and invariant mass $m_{JJ} \geq 1$ TeV. The presence of a signal can be detected as a bump in the dijet invariant mass distribution. We present in Fig.~\ref{fig:mJJ-Zp} this distribution for the relevant decays of the $Z'$ boson, in the benchmark with $m_{Z'} = 2$ TeV. Notice that the $Z' \to t \bar t$ contribution has a large tail at lower $m_{JJ}$, due to the semilptonic top decays that are included in our simulation. For the rest of signals, the peak is slightly shifted due to the jet grooming. Although the different signal components do not peak exactly at the same mass, the combination of the several contributions is still possible.

The sample of events passing these criteria is divided into six classes according to the number of $b$ tags on the two jets:
\begin{itemize}
\item[(i)] two double $b$ tags (2DT);
\item[(ii)] a double $b$ tag and a single $b$ tag (1D1ST);
\item[(iii)] a double $b$ tag (1DT);
\item[(iv)] two single $b$ tags (2ST);
\item[(v)] a single $b$ tag (1ST);
\item[(vi)] no $b$ tags (0T).
\end{itemize}
The background composition depends on $m_{JJ}$, with $t \bar t$ and $b \bar b$ decreasing faster than $jj$ because the latter includes processes with two valence quarks. At this level of event selection dijet production is overwhelmingly dominant. As a general rule, the higher the number of $b$ tags, the smaller this background is because of the mistag suppression. 

\begin{figure}[t]
\begin{center}
\begin{tabular}{c}
\includegraphics[height=5.2cm,clip=]{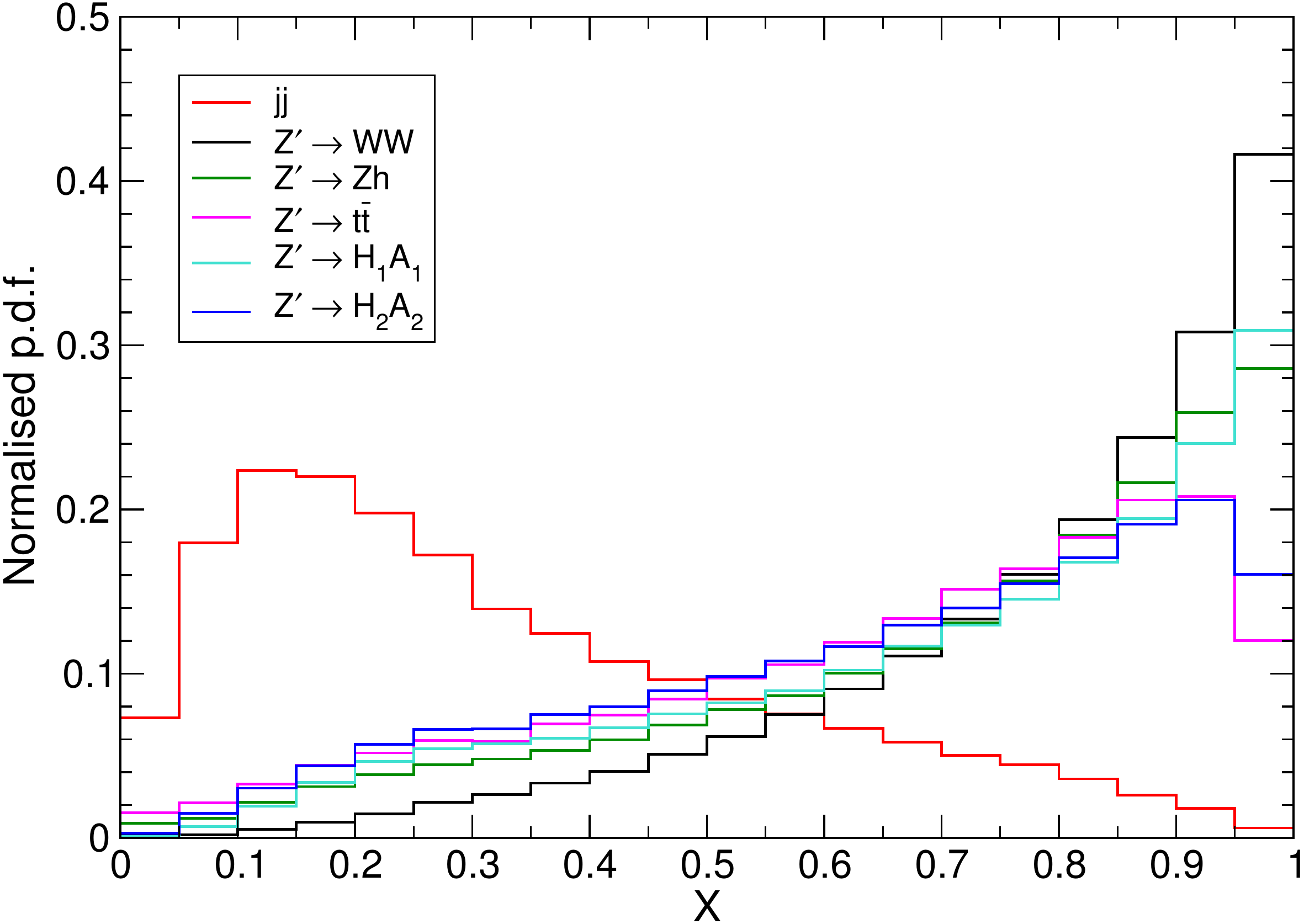}  
\end{tabular}
\caption{Left: Distribution of the NN score $X$ for the QCD background and various $Z'$ signals.}
\label{fig:NNscore}
\end{center}
\end{figure}

We use the MUST inspired tagger {\tt GenT} in Ref.~\cite{Aguilar-Saavedra:2020uhm} that classifies any type of multi-pronged jets as signal, and QCD jets as background. The NN score $X$ is shown in Fig.~\ref{fig:NNscore} for the QCD dijet background and the several signals arising from $Z'$ decays. (For each event with two jets, we use both jets in the plot.) We omit $Z' \to jj$ which is background-like and obviously suppressed by pre-selection cuts $m_J \geq 50$ GeV, and further by event selection. A dijet mass window $m_{JJ} \in [1.7,2.3]$ TeV is required, and the 2 TeV $Z'$ benchmark is used. As it is expected, jets stemming from the five $Z'$ decay modes into massive particles are correctly classified as signal with a distribution leaning to the r.h.s. of the plot, while QCD jets are classified as background. Note that with this tagger the continuum $t \bar t$ production is not rejected as background, as it produces multi-pronged jets like the $Z' \to t \bar t$ signal ones. 

\begin{figure}[t!]
\begin{center}
\begin{tabular}{cc}
\includegraphics[height=5.5cm,clip=]{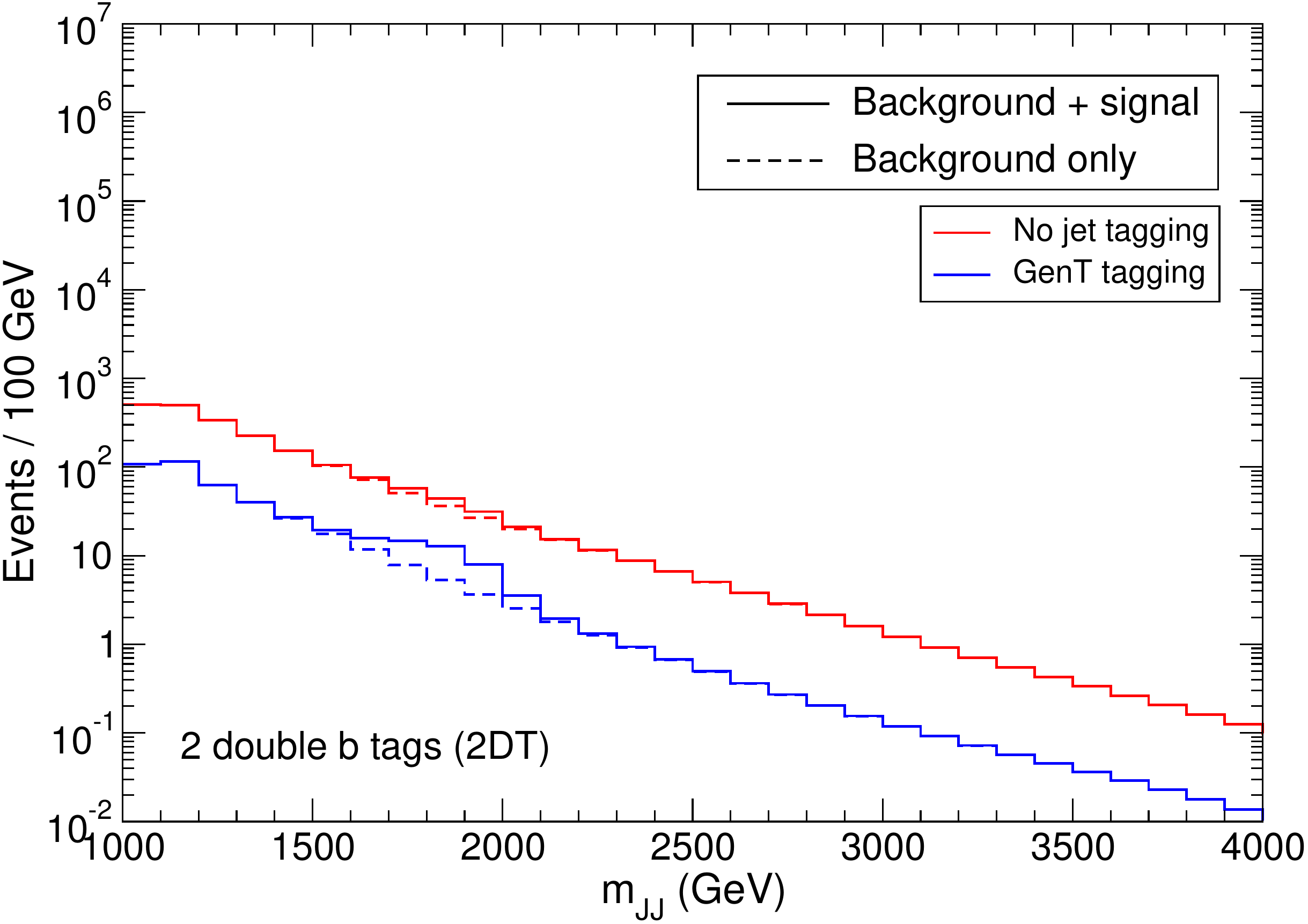} &
\includegraphics[height=5.5cm,clip=]{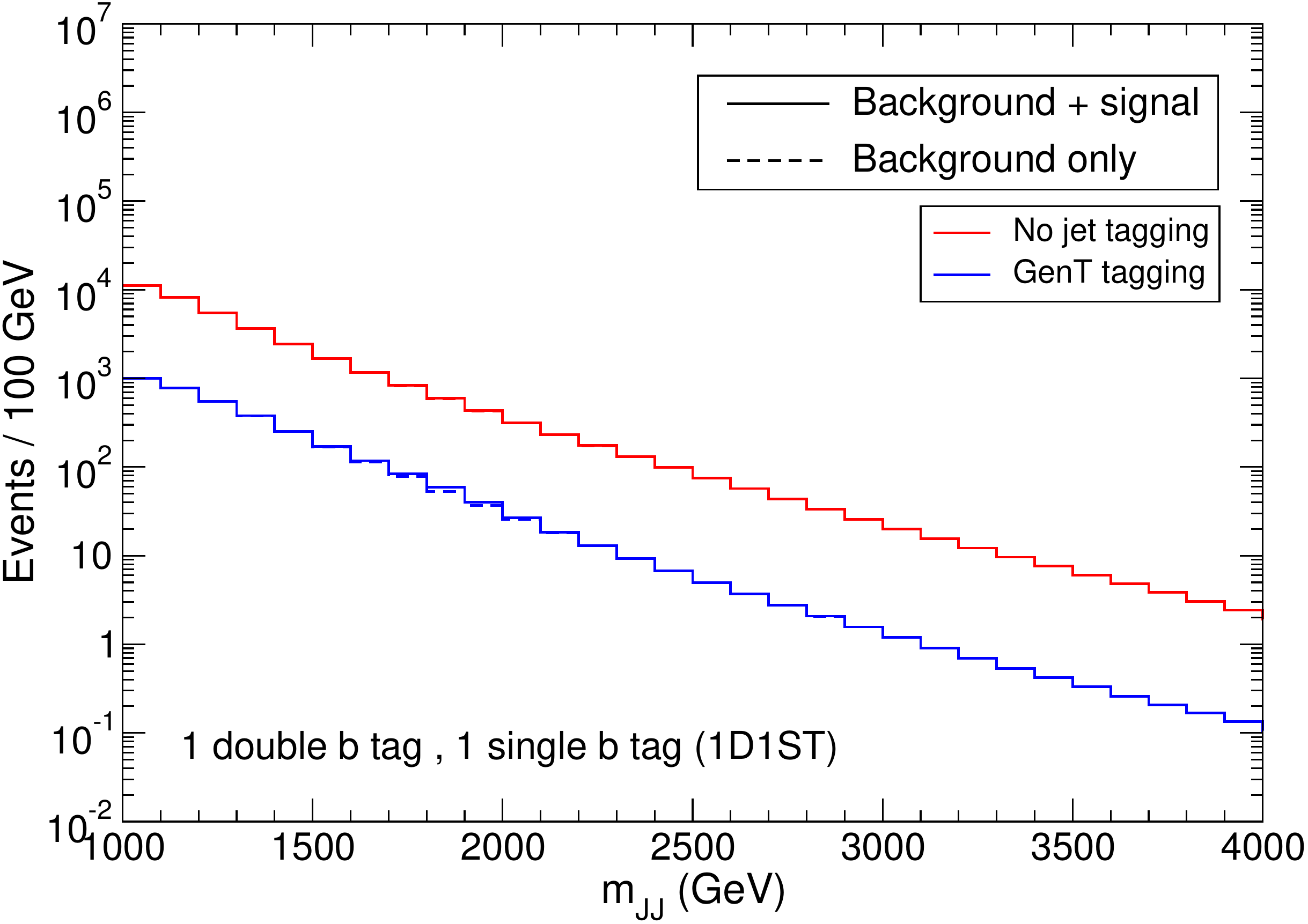}  \\
\includegraphics[height=5.5cm,clip=]{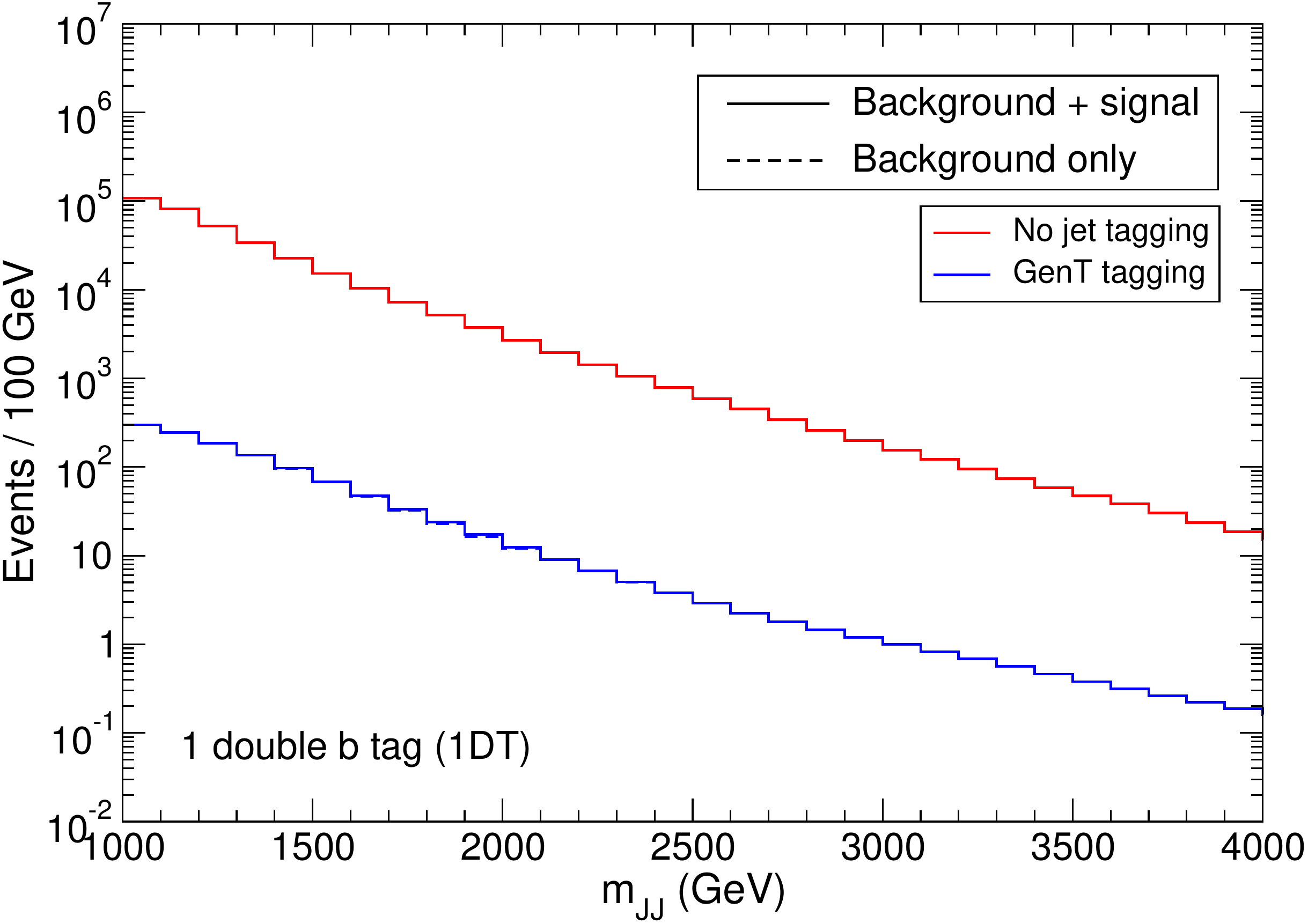}  &
\includegraphics[height=5.5cm,clip=]{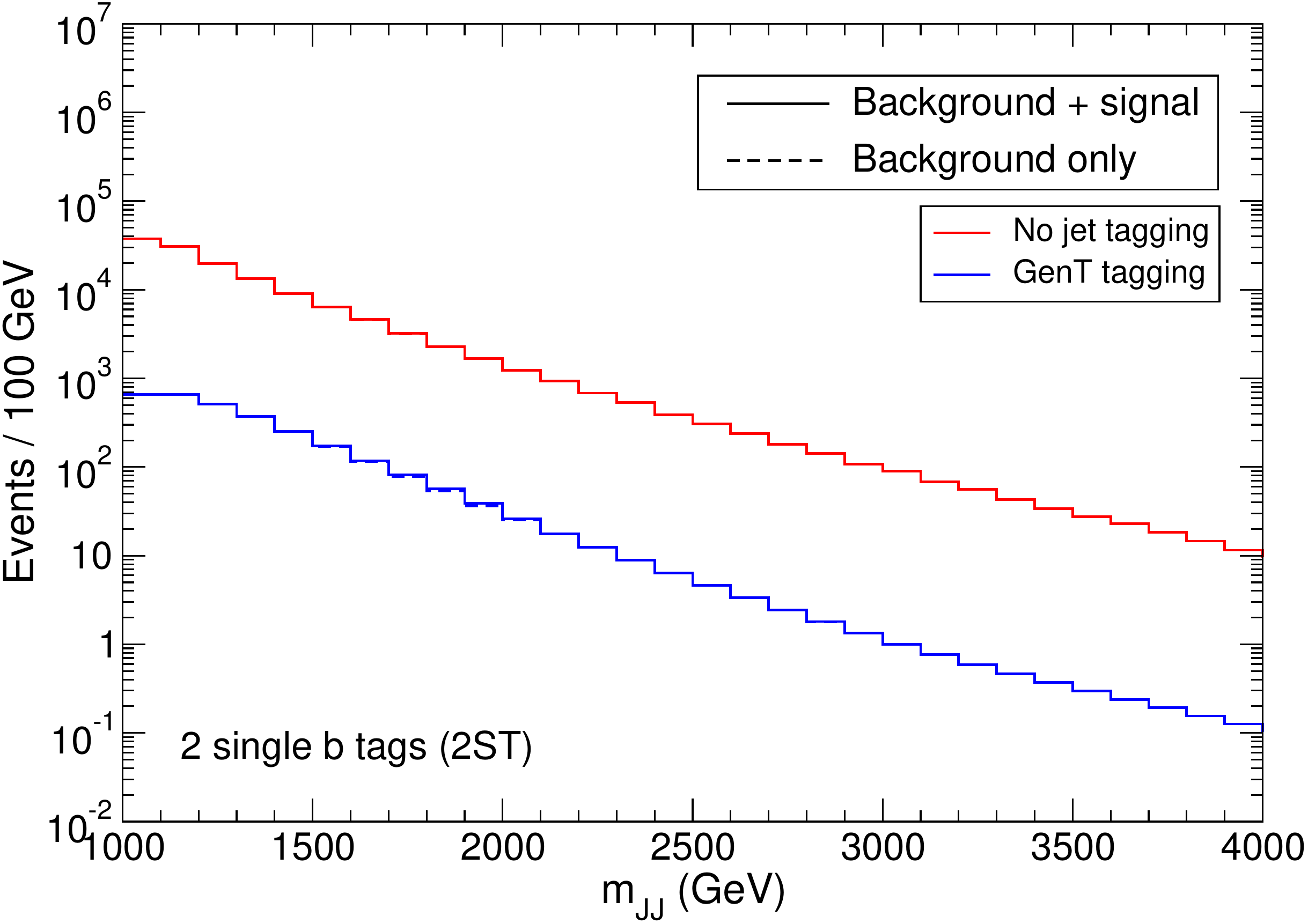} \\
\includegraphics[height=5.5cm,clip=]{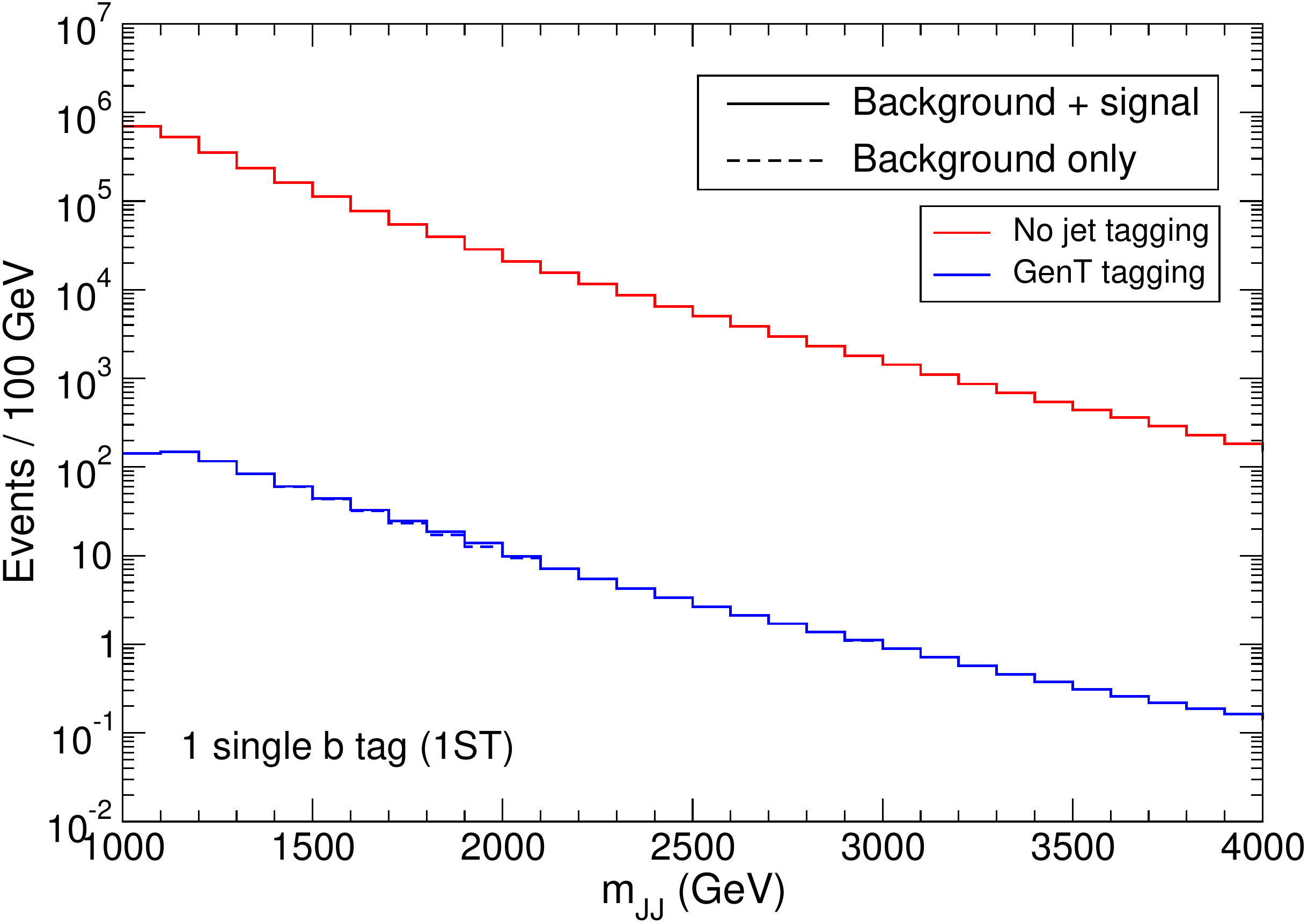}  &
\includegraphics[height=5.5cm,clip=]{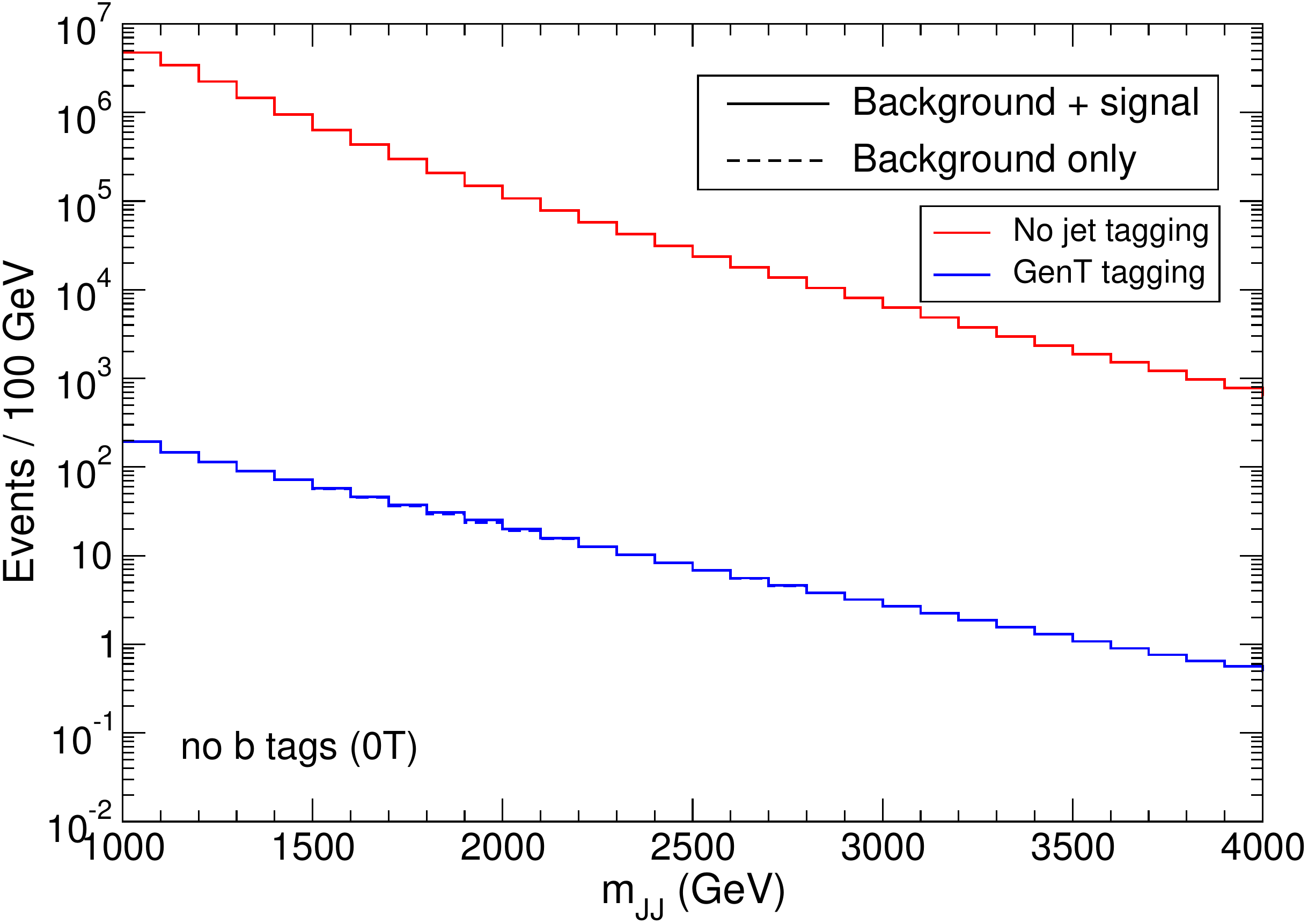} 
\end{tabular}
\caption{Invariant mass distribution for the background plus an injected 2 TeV $Z'$ signal in the six event classes, before and after {\tt GenT} tagging.}
\label{fig:dist}
\end{center}
\end{figure}

For the final event selection we require, in addition, a lower cut on the NN score for the two jets that reduces the huge dijet background. The precise value of the cut is not optimised. Rather, we set a generic cut for each sample, such that the background is reduced to $O(1)$ event per 100 GeV bin at $m_{JJ} \simeq 3$ TeV. Specifically, we require
\begin{itemize}
\item Sample 2DT: $X \geq 0.5$;
\item Sample 1D1ST: $X \geq 0.6$;
\item Samples 1DT, and 2ST: $X \geq 0.8$;
\item Samples 1ST and 0T: $X \geq 0.9$.
\end{itemize}
In the 2DT and 1D1ST samples the background is already small, therefore the cuts are rather mild. On the other hand, stringent cuts are required to reduce the background in the 1ST and 0T samples. The total background plus the $Z'$ signal is shown in Fig.~\ref{fig:dist} for the six event classes, before and after {\tt GenT} tagging. The luminosity is taken as $L = 139~\text{fb}^{-1}$. Only in the 2DT class the presence of the signal is noticeable by eye, as a bump on the falling distribution after {\tt GenT} tagging. For illustration, we collect in Table~\ref{tab:comp} the background composition in the region $m_{JJ} \in [1.7,2.3]$ TeV in the six event classes.

\begin{table}[htb]
\begin{center}
\begin{tabular}{|c|c|c|c|c|c|c|}
\hline
& 2DT & 1D1ST & 1DT & 2ST & 1ST & 0T
\\ \hline
$jj$            & 48\%  & 41\%  & 78\%   & 9\%   & 40\%  & 92\%
\\ \hline
$b \bar b$ & 17\%   & 6\%   & 1\%    & 1\%    & $< 0.5\%$  & $<0.1\%$
\\ \hline
$t \bar t$   &  35\%  & 53\%  & 21\%   & 90\%   & 60\%   & 8\% 
\\
\hline
\end{tabular}
\caption{Background composition for $m_{JJ} \in [1.7,2.3]$ TeV, for the six event classes.}
\label{tab:comp}
\end{center}
\end{table}

Note that in some classes the background is dominated by $t \bar t$, meaning that the $b$ quark and {\tt GenT} tagging strongly suppress dijet production. We have not attempted further suppression of the $t \bar t$ continuum background since there is a $Z' \to t \bar t$ signal too, especially contributing in the 2ST class. We prefer to keep an inclusive event selection to just reject dijet and $b \bar b$ production.

\begin{figure}[t]
\begin{center}
\begin{tabular}{cc}
\includegraphics[height=5.2cm,clip=]{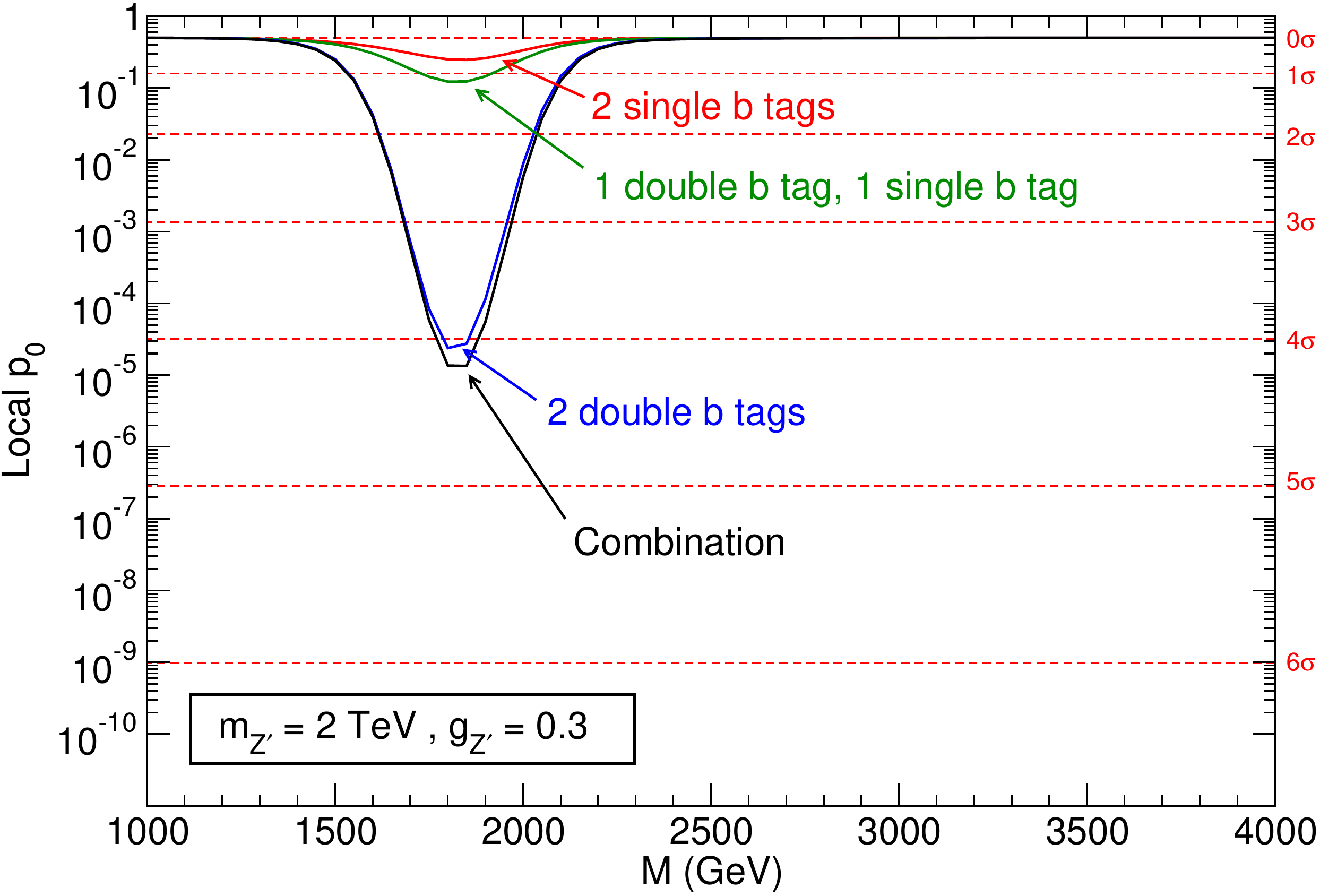} &
\includegraphics[height=5.2cm,clip=]{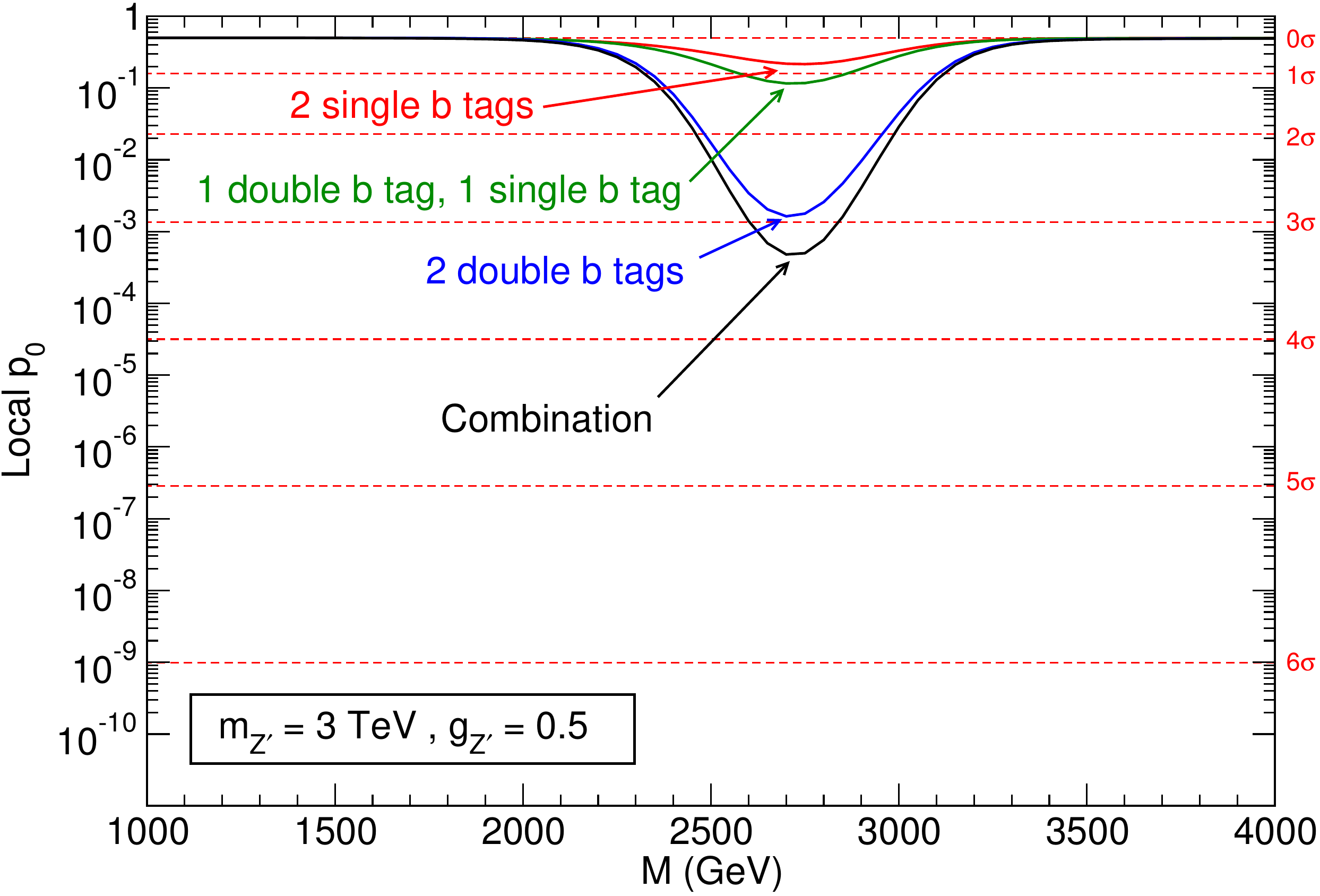}
\end{tabular}
\caption{Expected local $p$-value for the $Z'$ signal in the 6-channel combination, as well as in selected event classes.}
\label{fig:Pval}
\end{center}
\end{figure}

The expected significance of the $Z'$ signal in the different searches is computed by using the Monte Carlo predictions for signal plus background as pseudo-data, performing likelihood tests for the presence of narrow resonances over the expected background, using the $\text{CL}_\text{s}$ method~\cite{Read:2002hq} with the asymptotic approximation of Ref.~\cite{Cowan:2010js}, and computing the $p$-value corresponding to each hypothesis for the resonance mass. The probability density functions of the potential narrow resonance signals are Gaussians with centre $M$ (i.e. the resonance mass probed) and standard deviation of $0.05 M$ (of the order of the experimental resolution). 
The results are given in figure~\ref{fig:Pval}, for a luminosity of $139~\text{fb}^{-1}$. In the left panel we take $M_{Z'} = 2$ TeV with a coupling $g_{Z'} = 0.3$, and in the right panel we take $M_{Z'} = 3$ TeV, $g_{Z'} = 0.5$. These couplings are smaller than the 95\% CL upper limits from searches, which are $g_{Z'} \leq 0.38$ and $g_{Z'} \leq 0.69$ for these two $Z'$ masses, respectively.

The potential significance of the $Z'$ signals in this generic search is remarkable, bearing in mind the values of the coupling we have used as benchmark.\footnote{Notice that with the normalisation we have used, the $U(1)'$ charges for SM quarks are relatively small, therefore the $Z'$ production cross sections for these couplings are quite smaller than for other $Z'$ bosons common in the literature.} For $M_{Z'} = 2$ TeV, $g_{Z'} = 0.3$, we have $\sigma(pp \to Z' \to Zh) = 0.6$ fb, well below the current upper limit of 0.95 fb (see Fig.~\ref{fig:lim}). For $M_{Z'} = 3$ TeV, $g_{Z'} = 0.5$ we have 
$\sigma(pp \to Z' \to WW) = 0.16$ fb, to be compared to the limit of 0.30 fb in Fig.~\ref{fig:lim}. The sensitivity to the $Z'$ signals is driven by the 2DT class where the background can be made quite small with little reduction on the signal. To this class, $Z' \to H_2 A_2$ is the main contributing decay mode. In other event classes there are important contributions from $Z' \to H_2 A_2$, $Z' \to t \bar t$ and $Z' \to Zh$. The decay $Z' \to WW$ is only relevant for the 0T class.

Finally, let us comment that one could worry about the extra $Z'$ decay modes present in this model that might contribute to the signal in the searches targetting SM decay modes~\cite{Sirunyan:2021bzu,Aad:2019fbh}, which yield the strongent constraints on the $Z'$ coupling. For $Z' \to Zh$, which gives the most stringent limit for $M_{Z'} = 2$ TeV, the CMS Collaboration uses leptonic $Z$ boson decays in Ref.~\cite{Sirunyan:2021bzu}, so the new modes $Z' \to H_2 A_2$, $Z' \to H_1 A_1$ do not contribute. For $Z' \to WW$ one expects some contribution from $Z' \to H_1 A_1$ in the signal region. Nevertheless, the limit on $g_{Z'}$ from this channel is much looser for this $Z'$ mass. On the other hand, for $M_{Z'} = 3$ TeV the most stringent direct limit is from $Z' \to WW$. With the masses $M_{H_1,A_1} = 96$ GeV used, one expects some overlap with the jet mass window used in Ref.~\cite{Aad:2019fbh}. Nevertheless, the two-pronged jet tagging efficiencies are generically smaller for jets with $b$ quarks, and the coupling used gives a cross section that is substantially smaller than the upper limit from this search.

\section{Discussion}
\label{sec:4}

In this paper we have used as benchmark a $Z'$ model in which, in addition to decays into SM particles, $Z' \to Zh$, $Z' \to WW$, $Z \to t \bar t$, $Z' \to jj$, the new boson decays into new scalars $Z' \to H_i A_i$, $i=1,2$. The latter decay produces multi-pronged jets containing several $b$ quarks. The goal of this paper has been to outline a general search strategy that could enable to detect such decays.

The presence of $Z' \to Zh$, $Z' \to WW$ decays strongly constrains the model, that is, for a given $Z'$ mass, the non-observation of any excess in current searches~\cite{Sirunyan:2021bzu,Aad:2019fbh} sets strong constraints on the $Z'$ coupling, which in turn constrains the possible cross section for the new modes $\sigma (pp \to Z' \to H_i A_i)$ to be around 1.5 fb for $M_{Z'} = 2$ TeV, and 0.5 fb for $M_{Z'} = 3$ TeV, at most. In this quite constrained parameter space, we have seen that a generic search using
\begin{itemize}
\item[(1)] $b$ tagging of sub-jets to divide the event sample into different categories according to the number of $b$ tags;
\item[(2)] a generic jet substructure tagger that selects any type of multi-pronged jets,
\end{itemize}
is very effective, with an additional requirement of sizeable jet masses. (Here we have taken $m_J \geq 50$ GeV.) The benefit of this strategy is to be sensitive to the exotic decay modes of the $Z'$ boson that also produce two large-radius jets. In this context, it is worth pointing out that consistent models can be built in which the exotic $Z'$ decays into scalars are dominant~\cite{Aguilar-Saavedra:2019adu}, with branching ratio around 50\%. In such case, a general search strategy to encompass all possible jet topologies is absolutely necessary.

\begin{figure}[t!]
\begin{center}
\begin{tabular}{c}
\includegraphics[height=8cm,clip=]{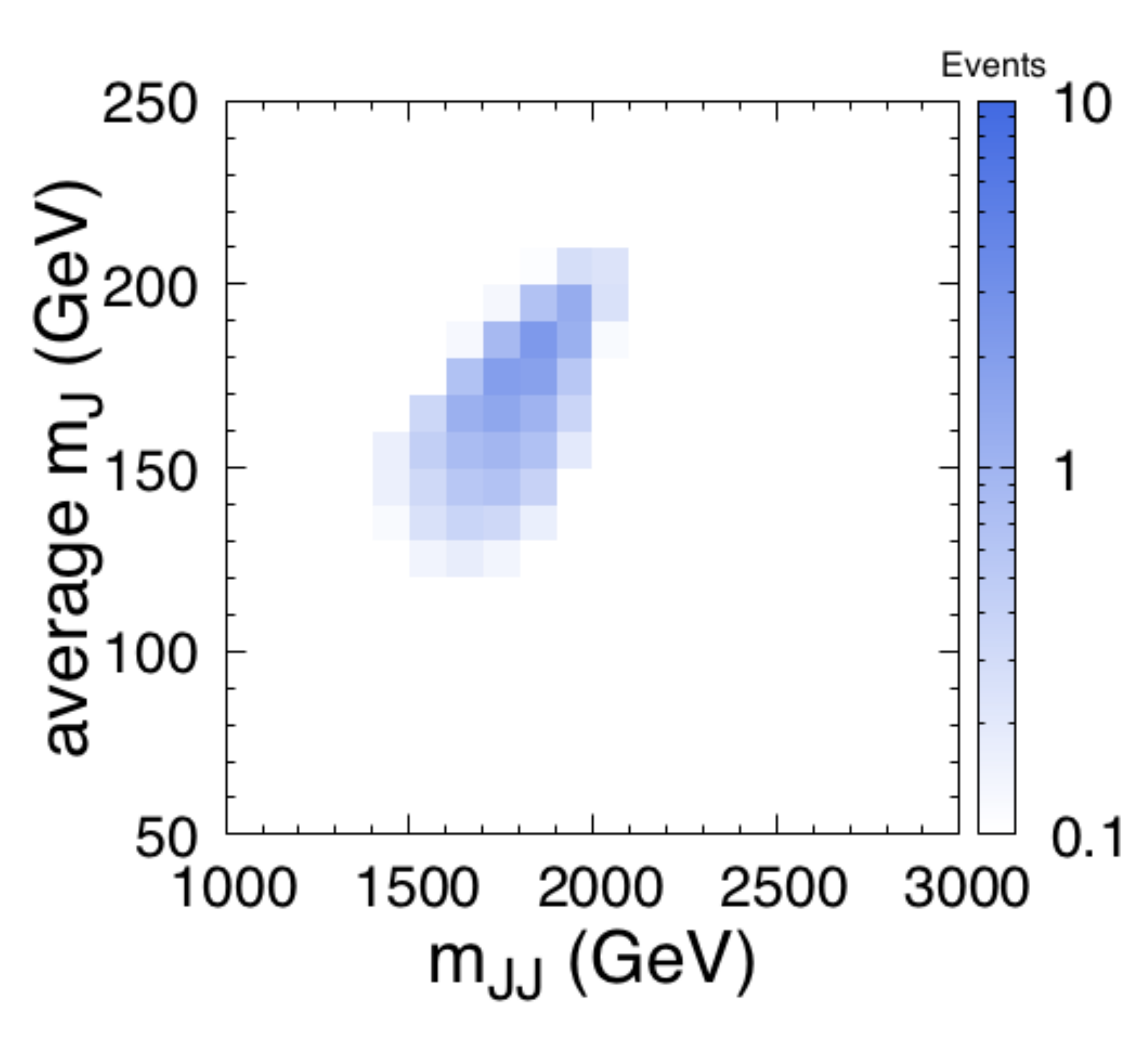}  
\end{tabular}
\caption{Two-dimensional distribution of the signal in the 2DT class, for the 2 TeV benchmark.}
\label{fig:2D}
\end{center}
\end{figure}

We remark that more sophisticated strategies can be pursued that might enhance the significance of a possible signal. In particular, we have selected a generic requirement $m_J \geq 50$ GeV while in several of the event classes the signal concentrates on a relatively narrow $m_J$ interval. For example, we show in Fig.~\ref{fig:2D} the distribution of signal events for the 2DT class (the most sensitive one) in the $m_{JJ}$, $m_J$ plane, where $m_J$ is the average of the two jet masses $m_{J_1}$, $m_{J_2}$ in the event. Clearly, in this specific case the background can be substantially reduced, without affecting the signal, by increasing the lower cut on $m_J$, and setting an upper cut. In an actual search, the optimisation can be done by considering a two-dimensional grid $(m_{JJ},m_J)$ or even a three-dimensional one $(m_{JJ},m_{J_1},m_{J_2})$. The CMS Collaboration has already performed a two-dimensional search~\cite{Sirunyan:2019jbg} for new resonances decaying into a pair of weak $W,Z$ bosons, using a tagging variable for two-pronged jets. Therefore, generic searches such as the one proposed in this paper, with a classification by categories based on $b$ tagging and a generic multi-pronged jet tagger, seem quite feasible and sensitive to new signals from heavy resonances.

\begin{acknowledgments}

We thank Javier Aguilar Saavedra for the use of computing resources.
The research of JAAS was supported by the Spanish Agencia Estatal de Investigaci\'on (AEI) through project PID2019-110058GB-C21.
The work of IL was funded by the Norwegian Financial Mechanism 2014-2021,
grant DEC-2019/34/H/ST2/00707.
The work 
of DL was supported by the Argentinian CONICET, and also acknowledges the support through PIP 11220170100154CO.
The research of CM was supported by the Spanish AEI 
through the grants 
PGC2018-095161-B-I00 and IFT Centro de Excelencia Severo Ochoa SEV-2016-0597.  
The authors acknowledge the support of the Spanish Red Consolider MultiDark FPA2017-90566-REDC.

\end{acknowledgments}

\end{document}